\documentclass[amsmath, amssymb, superscriptaddress, reprint, twocolumn, colorlinks=true, citecolor=blue, linkcolor=blue, urlcolor=blue]{revtex4-1}
\usepackage{natbib}
\usepackage[utf8]{inputenc}
\usepackage[T1]{fontenc}
\usepackage{graphicx}
\usepackage{grffile}
\usepackage{longtable}
\usepackage{wrapfig}
\usepackage{rotating}
\usepackage[normalem]{ulem}
\usepackage{amsmath}
\usepackage{textcomp}
\usepackage{amssymb}
\usepackage{capt-of}
\usepackage{hyperref}
\usepackage{mathptmx}
\usepackage{upgreek}
\usepackage{float}
\usepackage{bm}
\usepackage{booktabs}
\usepackage{subfigure}
\usepackage{url}
\usepackage{tablefootnote}

\date{\today; Authors to whom correspondence should be addressed: zihaofeng1998@163.com}

\begin{document}

\title{Computational Model for Photoionization in Pure SF\(_6\) Streamer at 1–15 atm}

\author{Zihao Feng*}
\affiliation{Department of Electrical Engineering, Tsinghua University, Beijing 100084, China}
\author{Liyang Zhang}
\affiliation{Department of Electrical Engineering, Tsinghua University, Beijing 100084, China}
\author{Xiaobing Zou}
\affiliation{Department of Electrical Engineering, Tsinghua University, Beijing 100084, China}
\author{Haiyun Luo}
\affiliation{Department of Electrical Engineering, Tsinghua University, Beijing 100084, China}

\begin{abstract}

Photoionization plays a crucial role in achieving accurate quantitative predictions in SF\(_6\) streamer simulations, but accurate models for SF\(_6\) photoionization remain limited, motivating this paper. First, we develop a computational model for SF\(_6\) photoionization and provide the detailed theoretical modeling process, as well as the comparison between experiment and simulation. A concise summary of model parameters within the comprehensive pressure range of 1 - 15 atm is provided for direct reference. Then, we perform comparative studies against simplified approaches. The results demonstrate that the proposed model effectively captures the non-local effects of SF\(_6\) photoionization, enhancing both the spatial numerical convergence and the accuracy of the streamer structure. Finally, we perform comparative studies by artificially increasing the photoionization intensity through multiplying the photoionization source term \(S_{\text{ph}}\) by a factor of 50 (\(50 \times S_{\text{ph}}\)) relative to the baseline intensity. Regarding breakdown voltage prediction, \(50 \times S_{\text{ph}}\) leads to a significant underestimate of the breakdown voltage for positive streamers, introducing errors greater than 0.5 kV, while exerting a small impact on negative streamers. Regarding streamer propagation dynamics, the radius of the positive streamer head exhibits pronounced shrinking, and \(50 \times S_{\text{ph}}\) reduces this shrinking and significantly lowers the head field by more than 700 Td. In contrast, \(50 \times S_{\text{ph}}\) has little impact on the morphology of the negative streamers and slightly enhances the head field by less than 30 Td.

\end{abstract}

\maketitle
\section{\label{1}introduction}

SF\(_6\) has been used as an insulating gas for nearly 70 years and remains the most widely used in gas-insulated electrical equipment \cite{8928273,https://doi.org/10.1002/tee.24244,10007898,10089517,10.1063/5.0295511,Li_2025}. Extensive experimental research on SF\(_6\) electrical discharges has observed some unusual discharge morphologies, and these insights are essential for understanding the nonlinear breakdown voltage behavior (\(U\)-\(P\) curve) of SF\(_6\) in non-uniform electric fields \cite{Waters_2019,4074858,Wu_2021,10.1088/1361-6463/add944,FPinnekamp_1983}. For instance, Gallimberti \textit{et al.} \cite{SF6111}, Seeger \textit{et al.} \cite{Seeger_2014},and Zhao \textit{et al.} \cite{Zhao_2022,https://doi.org/10.1049/hve2.12119} reported secondary streamers and leader precursors, which emerged from the SF\(_6\) streamer channel.  Wu \textit{et al.} \cite{Wu_2019} observed the positive glow corona within the SF\(_6\) streamer channel, whose shielding effect led to the occurrence of side sparks. Additionally, experimental and analytical research \cite{90289,Seeger_2009} has shown that SF\(_6\) streamer discharge plays a crucial role in initiating subsequent leader discharge and electrical breakdown. Therefore, a better understanding of the microscopic characteristics of SF\(_6\) streamer morphology and related breakdown voltage is of practical significance.

The plasma fluid simulations can resolve spatiotemporal microscopic characteristics of the streamer discharge \cite{Nijdam_2020}. In the context of fluid simulations of SF\(_6\)-related streamer, Morrow \cite{PhysRevA.35.1778} first reported the properties of SF\(_6\) streamers and streamer channels under uniform fields in 1987. According to an incomplete survey of the literature, many simulations have been conducted on streamers in pure SF\(_6\) \cite{14,GDYJ200807009,NYuBabaeva_2002,GDYJ201410015}, as well as in mixed gases such as SF\(_6\)/N\(_2\) \cite{10.1063/5.0006140,10400497,PhysRevA.37.4396} and SF\(_6\)/CO\(_2\) \cite{10.1063/5.0006140,10.1063/5.0076343,10230229}. More recently, Francisco \textit{et al.} \cite{Francisco_2021} identified general coherent structures-isolated streamer heads and ion conductive channels-in strong electronegative gases. Luo \textit{et al.} \cite{9099705} reported dynamics and chemical kinetics of SF\(_6\)/N\(_2\) streamers. Levko and Raja \cite{10.1063/5.0131780} investigated the breakdown voltage characteristics of SF\(_6\)/N\(_2\) streamers using a comprehensive plasma chemical reaction mechanism. Zhang \textit{et al.} \cite{10092907}, Wang \textit{et al.} \cite{Wang_2025}, and Zhong \textit{et al.} \cite{10969826} reported the dynamics of pure SF\(_6\) and SF\(_6\)/N\(_2\) streamers along dielectric surfaces. 

However, simulations of pure SF\(_6\) in highly non-uniform fields seem still limited. Based on some published results \cite{j5020018}, one may assume that the strong electronegativity of SF\(_6\) may cause difficulties in spatial numerical convergence. Specifically, after the streamer formed, strong electric field shielding occurs at the rear edge of the streamer head. The stronger the attachment of the gas, the more rapidly electrons are depleted in this position, leading to the formation of a region with a steep electron density gradient (\( \nabla n_\text{e} \)). Notably, in numerical simulations, this \( \nabla n_\text{e} \) poses challenges for spatial convergence numerically at the rear edge of the streamer head, but the non-local photoionization can supplement seed electrons to this region. In addressing this, Feng \textit{et al.} \cite{10.1063/5.0223522,arxiv} qualitatively modeled SF\(_6\) photoionization using simplified alternative approaches; however, these approaches were unrealistic and inadequate for quantitative studies or engineering predictions. Rose \textit{et al.} \cite{10.1063/1.3629989} employed an explicit kinetic approach to simulate SF\(_6\) photoionization in their particle-in-cell model, but some parameters, such as quenching pressure, need further estimation. Importantly, photo-radiation has been identified as a key parameter governing SF\(_6\) electrical discharge \cite{Seeger_2014,Seeger_2008123,Seeger_2009}. To date, a more accurate computational model for SF\(_6\) photoionization remains awaiting, which motivates the research presented in this paper.

In this paper, we develop a computational model for SF\(_6\) photoionization based on SF\(_6\)-related data. The model is built upon the theoretical framework of Zheleznyak's classical photoionization model \cite{ZK} and Pancheshnyi's analytical model \cite{Pancheshnyi_2015}, while the numerical computation follows the Helmholtz equation model proposed by Luque \textit{et al.} \cite{10.1063/1.2435934} (and in parallel by Bourdon \textit{et al.} \cite{Bourdon_2007}). A detailed description of the development of SF\(_6\) photoionization model is presented in Section \ref{2} and a comparison between experiment and simulation is presented in Appendix. B. Additionally, we conduct comparative studies of the photoionization model. In Section \ref{3.1}, we compare the proposed photoionization model with simplified alternative approaches, highlighting the crucial role of non-local photoionization in ensuring both spatial numerical convergence and accuracy in SF\(_6\) streamer simulations. In Section \ref{3.2}, we compare the artificially increased photoionization intensity with the baseline intensity, examining their effects on the accuracy of breakdown voltage predictions and the dynamics of SF\(_6\) streamer propagation. Section \ref{4} discusses the role of photoionization in streamer branching, as well as the applicability and limitations of 2D simulations. Finally, Section \ref{5} provides a concise summary of SF\(_6\) photoionization model and relating model parameters within the comprehensive pressure range of 1 - 15 atm for direct reference.

\section{\label{2}Development of SF\(_6\) Photoionization Model }

The theoretical foundation of model development is based on Zheleznyak's classical photoionization model. In Zheleznyak’s model \cite{ZK}, the photoionization source term \( S_{\text{ph}} \) at the observation point \(\boldsymbol{r}\) due to source points emitting UV photons at \( \boldsymbol{r}^{\prime}\) is defined as (with the distance between these two points \( |\mathbf{r} - \mathbf{r'}| = R \)): 

\begin{equation}
   \label{Eqphoto1}
S_{\mathrm{ph}}(\mathbf{r})=\int_{V_1} \frac{\sum_k I_k\left(\mathbf{r}^{\prime}\right) g_k({R})}{4 \pi R^2} \mathrm{~d} V_1
\end{equation}

, where \( g_k({R}) \) denotes the absorption function, and \( I_k(\boldsymbol{r}) \) denotes the photon production rate. \( k \) corresponds to the specific ionizing radiation for SF\(_6\), which is characterized by its wavelength in this paper.

\subsection{\label{2.1} Investigation of photon production rate}

The photon production rate for \(k\)-radiation \( I_{k}(\boldsymbol{r}) \) is defined as:

\begin{equation}
   \label{Eqphoto2}
{I}_{{k}}(\mathbf{r})=\frac{{p}_{\mathrm{q},{k}}}{{p}+{p}_{\mathrm{q},{k}}} {\xi}_{{k}} \frac{{v}_{\mathrm{u},{k}}}{{v}_{\mathrm{i}}} {S}_{\mathrm{i}}(\mathbf{r})
\end{equation}

, where \( \xi_k \) denotes the photoionization efficiency, \( v_{\mathrm{u},k} \) denotes the excitation frequency, \( v_{\mathrm{i}} \) denotes the ionization frequency, thus, \( \frac{v_{\mathrm{u},k}}{v_{\mathrm{i}}} \) represents the ionizing radiation efficiency. \( p_{\text{q},k} \) is the quenching pressure, reflecting the collision quenching (nonradiative deactivation) between excited states and neutral species. All the above correspond to \(k\)-radiation. \( S_{\mathrm{i}}(\boldsymbol{r}) = \alpha n_{\mathrm{e}} \left|\mu_{\mathrm{e}} \boldsymbol{E}\right| \) denotes the impact ionization rate, \(p\) denotes the gas pressure, which is set to 1 atm in the modeling process of this section; when determining the model parameters of the high-pressure conditions in Section \ref{5}, \(p\) is adjusted accordingly.

\begin{figure}[t]
\centering
\includegraphics[width=8.5cm]{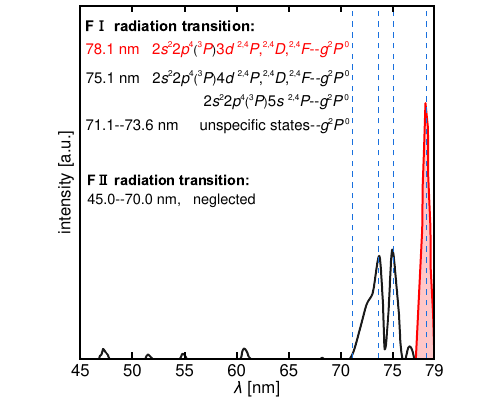}
\caption{\label{fig.spectrum} Further calibrated SF\(_6\) emission spectrum based on original spectral data of Forand \textit{et al.} \cite{doi:10.1139/p86-048}. The identification of different radiation transitions of F I is indicated.}
\end{figure}

\begin{figure}[b]
\centering
\includegraphics[width=8.5cm]{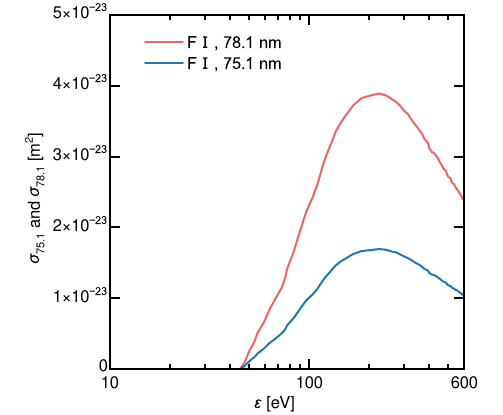}
\caption{\label{fig.section} Excitation functions for direct dissociative excitation from SF\(_6\) to F I 78.1 nm and F I 75.1 nm}
\end{figure}

To estimate these parameters, first, the calibrated SF\(_6\) emission spectrum measured by Forand \textit{et al.} \cite{doi:10.1139/p86-048} using 200 eV incident electrons is used to identify the excited states responsible for photoionization in pure SF\(_6\). Since the threshold ionization energy for SF\(_6\) is 15.7 eV, corresponding to a wavelength of 79 nm, we focus on the spectrum with \(\lambda < 79 \ \text{nm}\). The F II radiation are negligible compared to that of F I; therefore, only F I radiation is considered. However, Ref. \cite{doi:10.1139/p86-048} noted that their spectrum was obtained using a supersonic nozzle, leading to inaccurate relative line intensities. Therefore, we further calibrate the relative F I line intensities using their corresponding maximum excitation cross sections, i.e., absolute cross sections provided in Table 2 of Ref. \cite{doi:10.1139/p86-048}. This yields a further calibrated ratio of \(I_{78.1}\): \(I_{75.1}\): \(I_{73.6 }\)=1: 0.44: 0.38, as shown in Fig. \ref{fig.spectrum}; this further calibrated spectrum is used for quantitative analysis in this paper.

\noindent\hspace*{2em}\textbf{\textit{ Estimation of quenching pressure \( p_{\text{q},k} \).}} Quenching pressure is crucial in quantifying ionizing radiation, as described by Naidis \cite{Naidis_2006}. The quenching pressure of F I in pure SF\(_6\) is defined as:
\begin{equation}
   \label{Eqphoto5}
p_{\mathrm{q},k}=\frac{k_{\mathrm{B}} T_{\mathrm{g}}}{\tau_{0} k_{\mathrm{q}}}
\end{equation}
, where \(k_\text{B}\) denotes the Boltzmann's constant, \(T_\text{g}=300\text{K}\) denotes the gas temperature of streamer, \( k_\text{q} = 7.5 \times 10^{-10} \, \text{cm}^3 \, \text{s}^{-1} \), denotes the collision quenching rate of F I by SF\(_6\) molecules, taken from Ref. \cite{LRicheboeuf_1998}, and \( \tau_{0} \) denotes the lifetime of specific F I state.

According to Ref. \cite{CELIK2012566}, the lifetime of F I (3d) corresponding to 78.1 nm radiation is estimated \(\sim\)20 ns, and the lifetime of F I (4d) and F I (5s) corresponding to 75.1 nm radiation is estimated \(\sim\)70 ns. The 71.1–73.6 nm band has no identified electronic configurations in Ref. \cite{doi:10.1139/p86-048} or other literature; we attribute it to higher states, e.g., F I (5p,5d,6s,6d,7s), most of which have lifetimes of \(\sim\)200 ns. Thus, \(p_{\text{q},71.1-73.6}\) \(\approx\frac{1}{10}
\)\(p_{\text{q},78.1}\). Moreover, no excitation cross sections are reported for F I (5p,5d,6s,6d,7s), possibly due to the low excitation probabilities of $n \geq 5$ electronic configurations. 

In summary, we assume the contribution of the 71.1–73.6 nm radiation to the photoionization of SF\(_6\) is negligible, and $p_{\text{q},78.1} \approx 2\,\text{Torr}$, $p_{\text{q},75.1} \approx 0.6\,\text{Torr}$.

\noindent\hspace*{2em}\textbf{\textit{  Estimation of the ionizing radiation efficiency \( \frac{v_{\mathrm{u},k}}{v_{\mathrm{i}}} \).}} Chemical channels play an important role in quantifying plasma chemistry, as described by Ju and Starikovskiy \cite{Ju}. For a more realistic analysis, we include the influence of SF$_6$ collision dissociation on the chemical composition. 
Similar arguments were also reported by Pancheshnyi (see \textit{Section 5} of \cite{Pancheshnyi_2015}) and Li \textit{et al.} (see \textit{Section 2.3} of \cite{Li_2024}). 
As derived from Ref. \cite{10.1063/1.1288407} and tested by simulations, the ionizing radiation region is assumed to consist of 0.1\% F radicals and 99.9\% SF$_6$ molecules, a ratio adopted in all subsequent calculations. Since SF$_6$ exhibits strong photoabsorption for $\lambda < 110$ nm, trace radicals (F, SF$_5$, SF$_4$, SF$_3$, SF$_2$, SF, and S) contribute only to F I excitation rather than photoionization.

\begin{figure}[b]
\centering
\includegraphics[width=8.5cm]{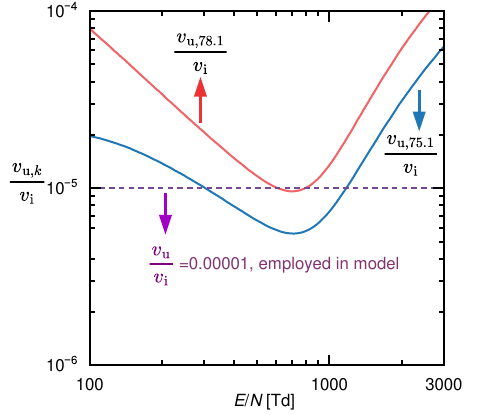}
\caption{\label{fig.efficient} Ionizing radiation efficiency of F I 78.1 nm and F I 75.1 nm as a function of the reduced electric field. The recommended value of $\frac{v_{\text{u}}}{v_{\mathrm{i}}}$ employed in photoionization model is indicated}
\end{figure}

Then, the electron collision cross sections (including elastic collision, 
impact ionization, and excitation) for F, taken from the BSR database \cite{BSR}, together with the excitation functions for direct dissociative excitation from SF\(_6\) to F I 78.1 nm and F I 75.1 nm, i.e., $\sigma_{78.1}$, $\sigma_{75.1}$ (see Fig. \ref{fig.section}) derived from Ref. \cite{doi:10.1139/p86-048}, are added into the reaction system in Appendix. A and input into Bolsig+ \cite{bosig+} to solve the 0D Boltzmann equation. Based on the Bolsig+ results, further calculations of \( \frac{v_{\mathrm{u},k}}{v_{\mathrm{i}}} \) are carried out, where only the F $\rightarrow$ F I (3d, 5s) reactions and the SF$_6 \rightarrow$ F I (78.1 nm, 75.1 nm) are considered for excitation contributions.  
The calculated \( \frac{v_{\text{u},78.1}}{v_{\mathrm{i}}} \) and \( \frac{v_{\text{u},75.1}}{v_{\mathrm{i}}} \) are presented in Fig. \ref{fig.efficient}, showing that ionizing radiation remains non-negligible across the entire electric field range, including the low-field region (e.g., within the streamer channel). To our knowledge, no cross-section data are available for F $\rightarrow$ FI (4d), leading to an underestimate of \( \frac{v_{\text{u},75.1}}{v_{\mathrm{i}}} \) value. This uncertainty can be corrected by modifying \( \frac{v_{\text{u},75.1}}{v_{\mathrm{i}}} \).

\begin{figure}[t]
\centering
\includegraphics[width=8.5cm]{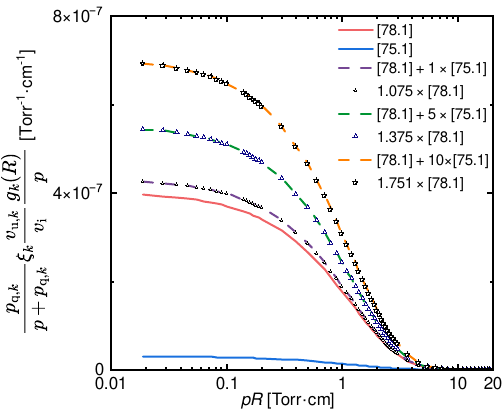}
\caption{\label{fig.zuhebianliang} Comparison of the calculated \(\frac{{p}_{\mathrm{q}, k}}{{p}+{p}_{\mathrm{q}, k}} \xi_k \frac{{v}_{\mathrm{u}, k}}{{v}_{\mathrm{i}}} \frac{{g}_k({R})}{{p}}\) term. In the legend, [78.1] denotes the term corresponding to the 78.1 nm radiation (\(k=78.1\)), and [75.1] denotes the term corresponding to the 75.1 nm radiation (\(k=75.1\)). The combined contribution is obtained by adding [78.1] to the [75.1] term multiplying by factors of 1×, 5×, 10×}
\end{figure}

As shown in Fig. \ref{fig.efficient}, the 75.1 nm radiation appears a secondary factor compared with the 78.1 nm radiation, but cannot be immediately neglected without verifying its influence on the profile of \(\frac{{p}_{\mathrm{q}}}{{p}+{p}_{\mathrm{q}}} \xi \frac{{v}_{\mathrm{u}}}{{v}_{\mathrm{i}}} \frac{{g}({R})}{{p}}\). Using Eq. \ref{Eqphoto10}, we first calculate the \(\frac{{p}_{\mathrm{q}, k}}{{p}+{p}_{\mathrm{q}, k}} \xi_k \frac{{v}_{\mathrm{u}, k}}{{v}_{\mathrm{i}}} \frac{{g}_k({R})}{{p}}\) corresponding to 75.1 nm radiation and 78.1 nm radiation. Then the 75.1 nm radiation term is corrected by multiplying 1×, 5×, 10× to represent the uncertainty caused by the lack of F $\rightarrow$ F I (4d) cross section, and subsequently added to the the 78.1 nm radiation term to obtain their combined contribution (fundamentally the overall \(\frac{{p}_{\mathrm{q}}}{{p}+{p}_{\mathrm{q}}} \xi \frac{{v}_{\mathrm{u}}}{{v}_{\mathrm{i}}} \frac{{g}({R})}{{p}}\)). These results are shown in Fig. \ref{fig.zuhebianliang} and demonstrate that the above procedure is equivalent to applying a multiplicative factor to the 78.1 nm contribution, indicating that the 75.1 nm radiation affects only the amplitude of the overall \(\frac{{p}_{\mathrm{q}}}{{p}+{p}_{\mathrm{q}}} \xi \frac{{v}_{\mathrm{u}}}{{v}_{\mathrm{i}}} \frac{{g}({R})}{{p}}\), not the profile. Thus, the contribution of 75.1 nm radiation to \(S_\text{ph}\) can be represented mathematically as a multiplicative factor \(M\) to \(\frac{{v}_{\mathrm{u},78.1}}{{v}_{\mathrm{i}}}\):

\begin{equation}
   \label{Eqphoto98}
\begin{aligned}
\begin{aligned}
& {S}_{\mathrm{ph}}(\mathbf{r})=\int_{V_1} \frac{{I}\left(\mathbf{r}^{\prime}\right) {g}({R})}{4 \pi {R}^2} \mathrm{~d} {V}_1 \\
& =\int_{V_1} \frac{\frac{{p}_{\mathrm{q}}}{{p}+{p}_{\mathrm{q}}} \xi \frac{{v}_{\mathrm{u}}}{{v}_{\mathrm{i}}} {S}_{\mathrm{i}}(\mathbf{r}^{\prime}) {g}({R})}{4 \pi {R}^2} \mathrm{~d} {V}_1 \\
& =\int_{V_1} \frac{\frac{{p}_{\mathrm{q}, 78.1}}{{p}+{p}_{\mathrm{q}, 78.1}} \xi_{78.1}\left({M} \times \frac{{v}_{\mathrm{u}, 78.1}}{{v}_{\mathrm{i}}}\right) {S}_{\mathrm{i}}(\mathbf{r}^{\prime}) {g}_{78.1}({R})}{4 \pi {R}^2} \mathrm{~d} {V}_1
\end{aligned}
\end{aligned}
\end{equation}

In summary, for pure SF$_6$ photoionization, it is sufficient to explicitly consider only the 78.1 nm radiation. The quantitative value of $\tfrac{v_{\text{u}}}{v_{\mathrm{i}}}=M\times\tfrac{v_{\text{u},78.1}}{v_{\mathrm{i}}}$ still requires estimation. To this end, we adjust $\tfrac{v_{\text{u}}}{v_{\mathrm{i}}}$ and conduct comparison between simulated and experimental positive streamer breakdown voltage. The rationale for employing the positive streamer breakdown voltage for comparison is because photoionization intensity strongly affects it, as discussed in Section \ref{3.2}. The comparison relies on (1) a single positive streamer dominates the breakdown; and (2) adjusting \( \frac{v_{\text{u}}}{v_{\mathrm{i}}} \) to match simulations with experiments. Details of the comparison procedure and experimental setup are provided in Appendix B. 

In summary, theoretical modeling and comparison between experiment and simulation recommend that the employed \( \frac{v_{\text{u}}}{v_{\mathrm{i}}} \) value in the classical fluid model under the local field approximation is estimated \( \frac{v_{\text{u}}}{v_{\mathrm{i}}}\approx 0.00001 \).

\begin{figure}[t]
\centering
\includegraphics[width=8.5cm]{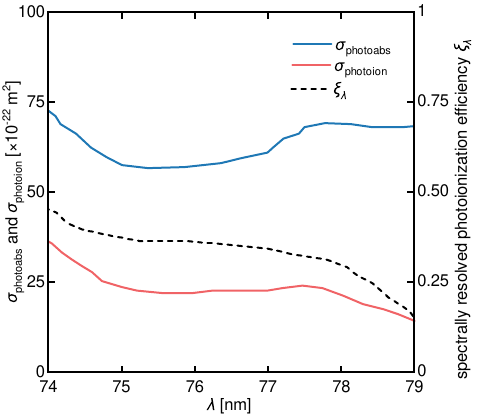}
\caption{\label{fig.reproduce} Reproduced data for SF\(_6\) spectrally resolved photoionization efficiency \(\xi_{\lambda}\) and SF\(_6\) photoionization cross section \(\sigma_{\text {photoion }}\) reported by Holland \textit{et al.} \cite{DMPHolland_1992}, SF\(_6\) photoabsorption cross section \(\sigma_{\text {photoabs }}\) reported by Ying \textit{et al.} \cite{10.1063/1.465149}. }
\end{figure}

\noindent\hspace*{2em}\textbf{\textit{ Estimation of photoionization efficiency \( \xi_k\).}} Since the above analysis confirms that only the contribution of 78.1 nm radiation needs to be explicitly considered, the overall photoionization efficiency $\xi=\xi_{78.1}$. Using the relative spectrum intensity in the range of 77.2–79 nm in Fig. \ref{fig.spectrum} as the weighting function, a weighted integral average of the spectrally resolved photoionization efficiency in Fig. \ref{fig.reproduce} is performed, yielding $\xi = \xi_{78.1} \approx 0.25$.

\subsection{\label{2.2} Investigation of absorption function}

To our knowledge, no experimental data are available for \(\frac{{g}({R})}{{p}}\) of pure SF\(_6\). Therefore, we refer to the analytical model proposed by Pancheshnyi \cite{Pancheshnyi_2015}, which in essence provided a formula for the photoionization function. In this paper, we follow Pancheshnyi’ s analytical model with a minor extension, and calculate the pressure-reduced absorption function \(\frac{{g}_{k}({R})}{{p}}\) integrated over wavelength. Specifically, by comparing the formula of Pancheshnyi \cite{Pancheshnyi_2015} and Zheleznyak \textit{et al.} \cite{ZK}, one can obtain:

\begin{equation}
   \label{Eqphoto10}
\frac{{g}_k({R})}{{p}}=\frac{1}{\xi_k} \cdot \frac{\int_{\lambda_{\min , k}}^{\lambda_{\max , k}} \xi_\lambda\left(\mu_\lambda / {p}\right) \exp \left(-\left(\mu_\lambda / {p}\right) {p} {R}\right) {I}_\lambda^0 \mathrm{~d} {\lambda}}{\int_{\lambda_{\min , k}}^{\lambda_{\max , k}} {I}_\lambda^0 \mathrm{~d} {\lambda}}
\end{equation}

, where \( (\lambda_{{\text{min},k}}, \lambda_{{\text{max},k}}) \) denotes the spectrum range for specific radiation, e.g., (77.2 nm, 79 nm) for 78.1 nm radiation. The \({I}_\lambda^0 \) denotes the spectral density in Fig. \ref{fig.spectrum}. The spectrally resolved photoionization efficiency \(\xi_\lambda\) is defined as:

\begin{equation}
   \label{Eqphoto11}
\xi_\lambda = \frac{\sigma_{\text {photoion }}(\lambda)}{\sigma_{\text {photoabs }}(\lambda)}
\end{equation}

, where \(\sigma_{\text {photoion }}\) denotes the SF\(_6\) photoionization cross section reported by Holland \textit{et al.} \cite{DMPHolland_1992}, and \(\sigma_{\text {photoabs }}\) denotes the SF\(_6\) photoabsorption cross section reported by Ying \textit{et al.} \cite{10.1063/1.465149}. All relevant data are reproduced in Fig. \ref{fig.reproduce}. The pressure-reduced spectrally resolved absorption coefficient \( \frac{\mu_\lambda}{p}\) is defined as:

\begin{equation}
   \label{Eqphoto12}
\frac{\mu_\lambda}{p} = \frac{\sigma_{\text {photoabs}}(\lambda)}{k_{\mathrm{B}} T_\text{g}}
\end{equation}

\begin{figure}[b]
\centering
\includegraphics[width=8.5cm]{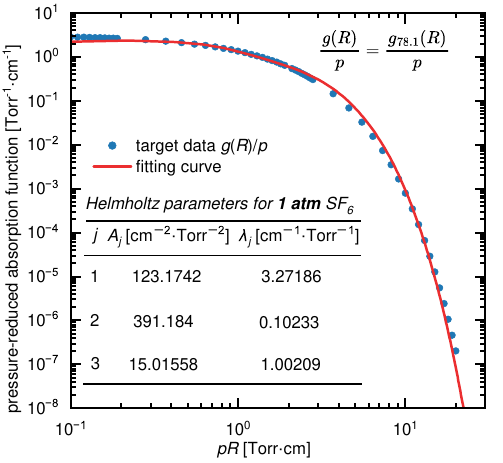}
\caption{\label{fig.fit} The calculated pressure-reduced absorption function \(\frac{g_{78.1}(R)}{p}\) (solid blue points), fitting curve (red line), and the fitting Helmholtz parameters are shown in the blank. All data corresponds to 1 atm SF\(_6\).}
\end{figure}

Finally, \( \frac{g_k(R)}{p} \) is solved using a calculation tool PHOTOPiC developed by Zhu \textit{et al.} \cite{zhu2020}, which was recently employed in Refs. \cite{zhu2021simulation,He_2024,Ma_2024,Kourtzanidis_2025}. The calculated result of \( \frac{g_{78.1}(R)}{p} \) is shown in Fig. \ref{fig.fit}.

\subsection{\label{2.3} Investigation of the Three-Term Helmholtz Equations}

In this paper, the Helmholtz equation model proposed by Luque \textit{et al.} \cite{10.1063/1.2435934} (and in parallel by Bourdon \textit{et al.} \cite{Bourdon_2007}) is employed for the numerical computation of the photoionization source term \(S_\text{ph}\). Its core idea is to approximate the pressure-reduced absorption function as follows:

\begin{equation}
   \label{Eqphoto6}
\frac{g(R)}{p} = (p R) \sum_j A_j \mathrm{e}^{-\lambda_j (p R)}
\end{equation}

This allows to transform the original integral equation (Eq.\ref{Eqphoto98}) into:

\begin{equation}
   \label{Eqphoto7}
S_{\mathrm{ph}}(\mathbf{r}) = \sum_j S_{\mathrm{ph}}^j(\mathbf{r})
\end{equation}

, with each term expressed as:

\begin{equation}
   \label{Eqphoto100}
S_{\mathrm{ph}}^j(\mathbf{r}) = \int_{V_1} \frac{I\left(\mathbf{r}^{\prime}\right) A_j p^2 \mathrm{e}^{-\lambda_j p R}}{4 \pi R} \mathrm{~d} V
\end{equation}

These terms satisfy the following Helmholtz partial differential equations (PDE):

\begin{equation}
   \label{Eqphoto9}
\nabla^2 S_{\mathrm{ph}}^j(\mathbf{r}) - \left(\lambda_j p\right)^2 S_{\mathrm{ph}}^j(\mathbf{r}) = -A_j p^2 I(\mathbf{r})
\end{equation}

This approach simplifies the original integral equation into a set of Helmholtz equations, thereby reducing computational costs. Here, after extensive testing to balance computational cost and accuracy, the three-term Helmholtz equations (for \( j = 1, 2, 3 \)) are chosen, with the parameters \( A_1,A_2, A_3, \lambda_1, \lambda_2, \lambda_3
 \) obtained by fitting Eq. \ref{Eqphoto6}.
 Specifically, the right-hand side of Eq. \ref{Eqphoto6} consists of three exponential terms with undetermined parameters (for \(j\) = 1,2,3), which are fitted using the Nelder-Mead simplex direct search method, while on the left-hand side of Eq. \ref{Eqphoto6} is the target data of \(\frac{g(R)}{p}\). The target data of $\frac{g(R)}{p} = \frac{g_{78.1}(R)}{p}$, fitting results and fitting parameters for 1 atm condition are shown in Fig. \ref{fig.fit}.

\begin{figure}[b]
\centering
\includegraphics[width=8.5cm]{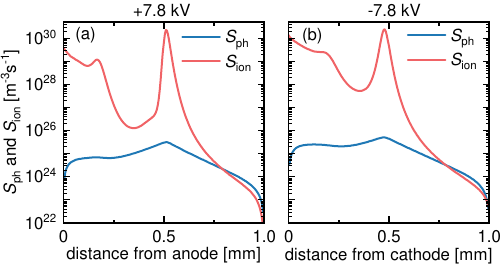}
\caption{\label{fig.com} Axial profiles of impact ionization rate \(S_{\text{ion}}\) and the photoionization rate \(S_{\text{ph}}\) for (a) \(U_0=+7.8\ \text{kV}\), and (b) \(U_0=-7.8 \ \text{kV}\).}
\end{figure}

\begin{figure*}[t]
\centering
\includegraphics[width=17cm]{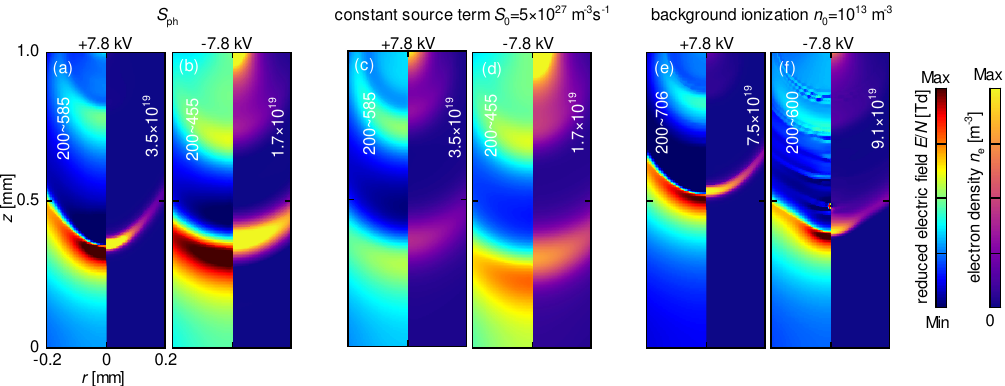}
\caption{\label{fig.model} The reduced electric field \(E/N\) and electron density \(n_\text{e}\) under different photoionization models and conditions: (a) \( S_{\text{ph}} \), \(U_0=+7.8 \ \text{kV}\); (b) \( S_{\text{ph}} \), \(U_0=-7.8 \ \text{kV}\); (c) \( S_{\text{0}} \), \(U_0=+7.8 \ \text{kV}\); (d) \( S_{\text{0}} \), \(U_0=-7.8 \ \text{kV}\); (e) \( n_{\text{0}} \), \(U_0=+7.8 \ \text{kV}\); (f) \( n_{\text{0}} \), \(U_0=-7.8 \ \text{kV}\).  Labels for \((E/N)_\text{min}\), \((E/N)_\text{max}\) and \(n_\text{e,max}\) are shown in each sub-figure, where \(n_\text{e,min}\) is fixed at 0.}
\end{figure*}

\subsection{\label{2.4}Evaluation of Non-local Effects of Proposed Photoionization Model}
We evaluate the proposed photoionization model using the 2D axisymmetric fluid model under the local field approximation in highly non-uniform fields created by the rod-plane electrode. The detailed description of numerical scheme is provided in Appendix A.

As shown in Fig. \ref{fig.com}(a) and \ref{fig.com}(b), for both positive and negative streamers, the photoionization rate (\(S_{\text{ph}}\)) reaches its peak near the streamer head, where it is \(\sim\)0.001\% of the maximum impact ionization rate (\(S_{\text{ion}}\)). Besides, the lowest values of \(S_{\text{ph}}\) and \(S_{\text{ion}}\) are both found within the streamer channel. Notably, a remarkable feature is observed: at the rear edge of the streamer head, the shielding effect for the electric field becomes significantly stronger, leading to a steep drop in \(S_{\text{ion}}\) by more than three orders of magnitude. But the \(S_{\text{ph}}\) is not directly affected by this localized suppression of ionization. Instead, \(S_{\text{ph}}\) exhibits a smooth spatial variation—by less than one order of magnitude—reflecting its non-local dependence on the overall ionizing radiation. Specifically, photoionization integrates photon contributions from ionizing radiation sources near the streamer head, and this effect contributes to the spatial numeral convergency as discussed in Section \ref{3.1}.

\section{\label{3}Comparative Study}
\subsection{\label{3.1}Comparative Study of Photoionization Models}

In the context of commonly used simplified alternative approaches for photoionization in pure SF\(_6\) or SF\(_6\) mixtures, some paper used the uniform background ionization, while some other added a constant source term to the continuity equation. The following discussions present a comparative study between the proposed photoionization model and the alternative approaches aforementioned. For comparison, the photoionization source term is set to \( S_{\text{ph}} \), a constant source term \( S_0 = 5 \times 10^{27}~\text{m}^{-3}\cdot\text{s}^{-1} \), and a uniform initial electron density \( n_0 = 10^{13}~\text{m}^{-3} \) (background ionization), respectively.

The constant source term \(S_0\) is uniformly applied across the computational domain, it results in a more diffuse space charge distribution near the streamer head. As a result, the electron density and electric field profiles in the \(S_0\) case (see Fig. \ref{fig.model}(c) and \ref{fig.model}(d)) become broader, with lower peak magnitudes, but extend over a larger region, including the streamer channel, compared to the \(S_{\text{ph}}\) case (see Fig. \ref{fig.model}(a) and \ref{fig.model}(b)). This deviation causes the pure SF\(_6\) streamer in the \(S_0\) case to not fully reflect the typical coherent structures, such as ion-conductive channels and isolated streamer heads \cite{Francisco_2021}.

In the \(n_0\) case, as shown in Fig. \ref{fig.model}(e) and \ref{fig.model}(f), even with a minimum grid spacing of 0.1 \(\mu\text{m}\), we failed to achieve well-converged numerical results, and the calculation was terminated after the streamer propagated only a short distance. This behavior resembles the situation described in Ref. \cite{j5020018}, where background ionization was also used to model photoionization. This is because the \(n_0\) case cannot provide sufficient seed electrons at positions with a steep \( \nabla n_\text{e} \), particularly at the rear edge of the streamer head, making spatial convergence at these positions more difficult. In contrast, the \(S_{\text{ph}}\) case could mitigate this issue, as already discussed in Section \ref{2.4}.

These results suggest that the non-local photoionization model presented in this paper, which supplies electrons both at the front of the streamer and at the rear edge of the streamer head, plays a crucial role in ensuring both spatial numerical convergence and accuracy. Also, it is important to distinguish that photoionization has a strong effect on the numerical convergence process, but not necessarily on the properties of the converged PDE solution itself.

\subsection{\label{3.2}Comparative Study of Photoionization Intensity}

\begin{figure*}[t]
\centering
\includegraphics[width=17cm]{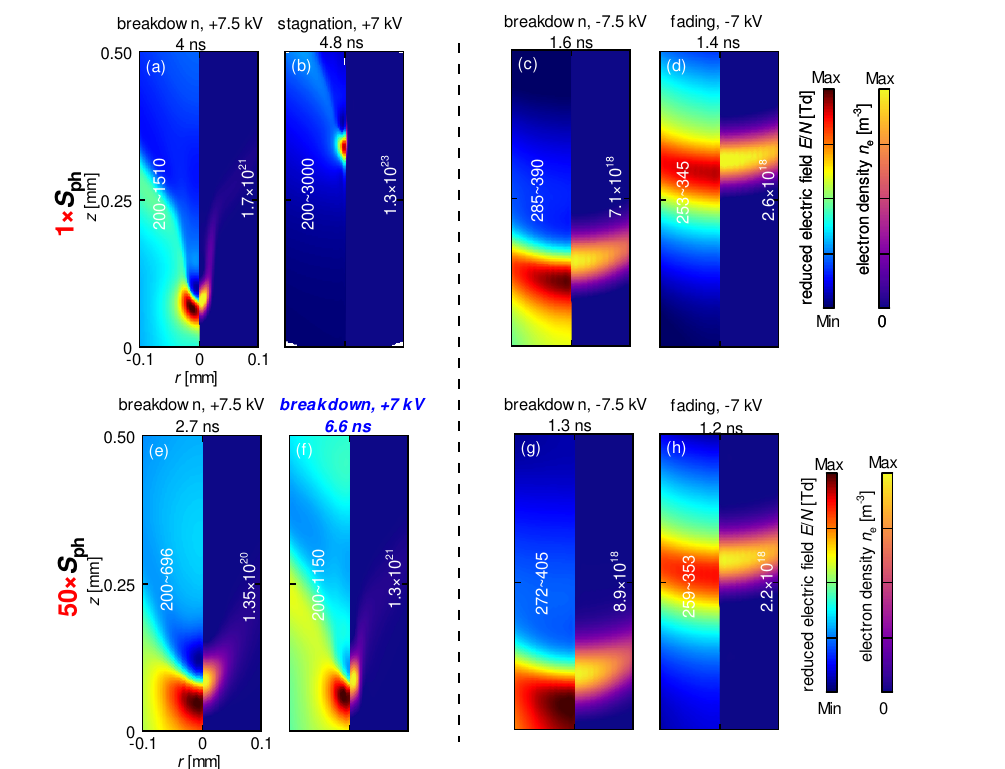}
\caption{\label{fig.volt} The reduced electric field \(E/N\) and electron density \(n_\text{e}\) under different photoionization intensity and conditions: (a) \( 1 \times S_{\text{ph}} \), \(U_0=+7.5 \ \text{kV}\); (b) \( 1 \times S_{\text{ph}} \), \(U_0=+7 \ \text{kV}\); (c) \( 1 \times S_{\text{ph}} \), \(U_0=-7.5 \ \text{kV}\); (d) \( 1 \times S_{\text{ph}} \), \(U_0=-7 \ \text{kV}\); (e) \( 50 \times S_{\text{ph}} \), \(U_0=+7.5 \ \text{kV}\); (f) \( 50 \times S_{\text{ph}} \), \(U_0=+7 \ \text{kV}\); (g) \( 50 \times S_{\text{ph}} \), \(U_0=-7.5 \ \text{kV}\); (h) \( 50 \times S_{\text{ph}} \), \(U_0=-7 \ \text{kV}\).  Labels for \((E/N)_\text{min}\), \((E/N)_\text{max}\) and \(n_\text{e,max}\) are shown in each sub-figure, where \(n_\text{e,min}\) is fixed at 0.}
\end{figure*}

As discussed in Section \ref{3.1}, seed electrons from photoionization are crucial for enhancing the spatial convergence of SF\(_6\) streamer simulations. This could inspire a possible idea that, for broader engineering predictions, increasing photoionization intensity might reduce computational costs by requiring fewer grids and resources. However, this assumption needs careful evaluation to assess its impact on the accuracy of breakdown voltage predictions and streamer propagation dynamics. The following two sub-sections provide an analysis of these two factors. For comparison, we modify the intensity of photoionization by adjusting the multiplier of the photoionization source term \(S_{\text{ph}}\) in the continuity equation, specifically setting it to \(1 \times S_{\text{ph}}\) (baseline) and \(50 \times S_{\text{ph}}\) (artificially increased).

\subsubsection{\label{3.2.1} Streamer Breakdown Voltage}

\begin{figure*}[t]
\centering
\includegraphics[width=17cm]{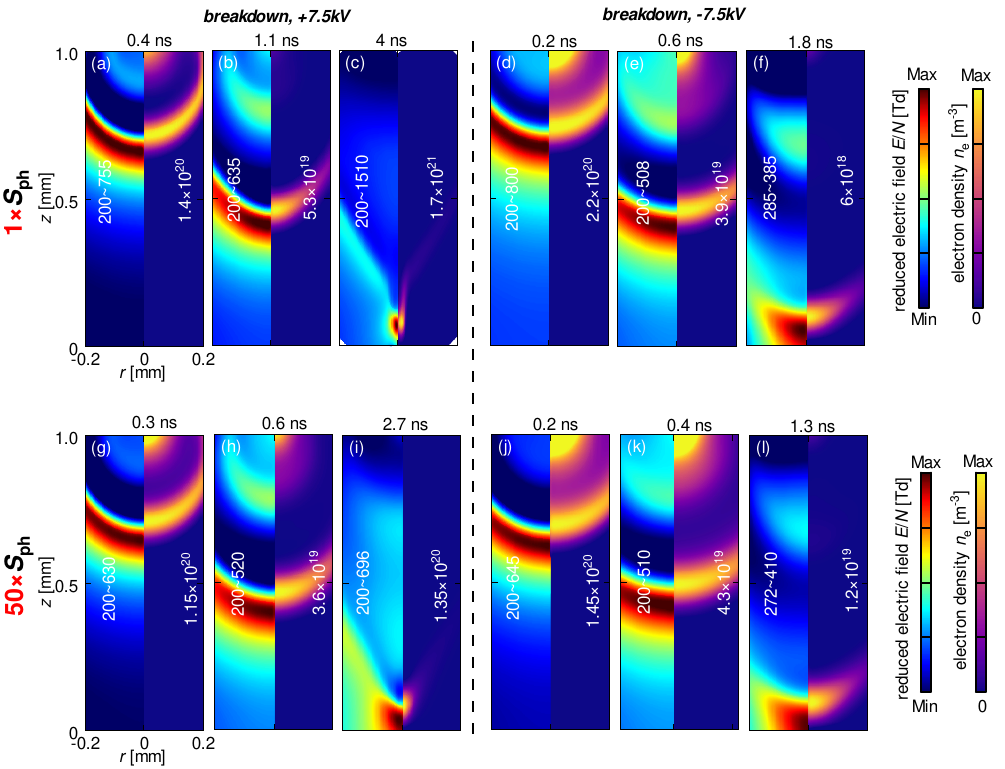}
\caption{\label{fig.sp} Evaluation of the reduced electric field \(E/N\) and electron density \(n_\text{e}\) under different photoionization intensity and conditions: (a-c) \( 1 \times S_{\text{ph}} \), \(U_0=+7.5 \ \text{kV}\); (d-f) \( 1 \times S_{\text{ph}} \), \(U_0=-7.5 \ \text{kV}\); (g-i) \( 50 \times S_{\text{ph}} \), \(U_0=+7.5 \ \text{kV}\); (j-l) \( 50 \times S_{\text{ph}} \), \(U_0=-7.5 \ \text{kV}\).  Labels for \((E/N)_\text{min}\), \((E/N)_\text{max}\) and \(n_\text{e,max}\) are shown in each sub-figure, where \(n_\text{e,min}\) is fixed at 0.}
\end{figure*}

In the simulation, breakdown voltage corresponds to simulation scenario where the streamer crosses the rod-plane gap, because it is considered to result in a conductive plasma channel that bridges the cathode and anode. Breakdown voltage error corresponds to simulation scenario where, if the applied voltage is set to the breakdown voltage minus this error, the streamer does not cross the gap.

For a single negative streamer event, when the electric field at the streamer head is below the threshold \((E/N)_{\text{cr}} = 360 \, \text{Td}\), the streamer is considered to be in a fading state, and breakdown does not occur, as reported in Refs. \cite{Guo_2022123,BKDEA}. In the \(1 \times S_{\text{ph}}\) case, the negative streamer can cross the entire gap at -7.5 kV (see Fig. \ref{fig.volt}(c)), while at -7 kV, fading occurs (see Fig. \ref{fig.volt}(d)). Setting the breakdown voltage error to 0.5 kV, the breakdown voltage is identified as -7.5 kV. In the \(50 \times S_{\text{ph}}\) case, the breakdown behavior of the streamer  (see Fig. \ref{fig.volt}(g) and \ref{fig.volt}(h)) is similar to that observed with \(1 \times S_{\text{ph}}\), with the breakdown voltage still identified as -7.5 kV. This indicates that artificially increasing the photoionization intensity to \(50 \times S_{\text{ph}}\) does not affect the accuracy of the breakdown voltage prediction for negative streamers.

For a single positive streamer event, the non-breakdown case is shown in Fig. \ref{fig.volt}(b). In this case, the head field rapidly increases from \(\sim\)1500 Td to \(\sim\)3000 Td within 0.1 ns, which is beyond the applicability of the drift-diffusion approximation. The electron density reaches \(10^{23} \, \text{m}^{-3}\), exceeding the electron density typical of a non-equilibrium, weakly ionized plasma. According to Refs. \cite{BKDEA,Niknezhad_2021,Li_2022}, this phenomenon is non-physical and results from the local source term expression used in the classical fluid model under local field approximation. Although it does not reflect the realistic distribution of electric field and electron density, it suggests that the positive streamer has stopped developing under this condition. This behavior is considered to be in a positive streamer stagnation state, and breakdown does not occur. As shown in Fig. \ref{fig.volt}(a) and \ref{fig.volt}(b), for \(1 \times S_{\text{ph}}\), when the breakdown voltage error is set at 0.5 kV, the breakdown voltage is identified as +7.5 kV. However, for \(50 \times S_{\text{ph}}\), the prediction of breakdown voltage shows a notable deviation. As shown in Fig. \ref{fig.volt}(e) and \ref{fig.volt}(f), the positive streamer can still break down at both +7 kV and +7.5 kV. This demonstrates that increasing the photoionization intensity by a factor of 50× significantly impacts the breakdown voltage prediction for positive streamers, causing a deviation greater than 0.5 kV and reducing the accuracy of the breakdown voltage prediction.

In summary, increasing the photoionization intensity to \(50 \times S_{\text{ph}}\) does not impact the breakdown voltage prediction for negative streamers, but it results in inaccurate predictions for positive streamers.

\subsubsection{\label{3.2.2}Streamer Propagation Dynamics}

\noindent\hspace*{2em}\textbf{\textit{Positive Streamer Shrinking}} We explain the SF\(_6\) positive streamer shrinking (see Figs. \ref{fig.sp}(c) and \ref{fig.sp}(i)) from the perspective of negative ion dynamics. This shrinking phenomenon is crucial for the subsequent comparative analysis. Similar shrinking was also observed in positive streamers in air, and Starikovskiy and collaborators \cite{10.1063/5.0037669,SVPancheshnyi_2004,Starikovskiy_2022} provided impressive explanations from the perspective of the electrical parameter. Here, we also compare our analysis with the core arguments of Refs. \cite{10.1063/5.0037669,SVPancheshnyi_2004,Starikovskiy_2022}, showing that these two perspectives are fundamentally equivalent. Specifically, in highly non-uniform fields, as positive streamer propagates away from the protrusion, the weakening background field strengthens attachment at the rear edge of the streamer head, generating more negative ions there. The accumulation of these negative ions corresponds to the loss of conductivity $\sigma$ (with $\sigma\downarrow \propto n_e\downarrow$) in Refs. \cite{10.1063/5.0037669,SVPancheshnyi_2004,Starikovskiy_2022}. These negative ions, in turn, modify the electric field: their space-charge field reduces the electric field on the axis and far from the axis. Due to the lower baseline electric field at positions farther from the axis, the total electric field at these positions drops below the effective ionization threshold (e.g., 360 Td for SF\(_6\)), causing the ionization wave profile to shrink toward the axis. This process corresponds to positive streamer head shrinking from the perspective of negative ion dynamics. After shrinking, the resulting space-charge density increases, causing the head field $E_{\text{head}}$ to rise ($E_{\text{head}}\uparrow$). In the electrical-parameter model, following Refs. \cite{10.1063/5.0037669,SVPancheshnyi_2004,Starikovskiy_2022}, the loss of conductivity ($\sigma\downarrow$) increases the potential drop across the channel, which in turn decreases the streamer head potential ($U_{\text{head}}\downarrow$). Thus, one can also conclude that the positive streamer-head radius \(R_{\text{head}}\) must shrink, with \(\
R_{\text{head}}\downarrow \ \propto\ \frac{U_{\text{head}}\downarrow}{E_{\text{head}}\uparrow}\, 
\), to ensure self-consistent solutions.

The extent of this shrinking depends on the distribution of background field, and strength of the attachment, with SF\(_6\) positive streamers showing more pronounced shrinking than air streamers \cite{Luque_2008,Nudnova_2008,SVPancheshnyi_2001,Bouwman_2025,10.1063/5.0292765}. Notably, for a positive streamer in 1 atm SF$_6$, simulations yield a post-shrinking head radius of $\sim$20 \(\mu\)m (see Fig. \ref{fig.sp}(c)), in agreement with experimental measurements (see Fig. 7 in Ref. \cite{Bujotzek_2015}), further confirming the validity of proposed photoionization model of this paper.

\noindent\hspace*{2em}\textbf{\textit{Head Profile}} The \(50 \times S_{\text{ph}}\) case has a notable effect on the profile of the positive streamer head. In the \(50 \times S_{\text{ph}}\) case, although the streamer head still shrinks after a period of development, its radius becomes larger (40 \(\mu \text{m}\), see Fig. \ref{fig.sp}(i)) compared to the \(1 \times S_{\text{ph}}\) case (20 \(\mu \text{m}\), see Fig. \ref{fig.sp}(c)), and the electron density profile becomes more diffuse. This is primarily because photoionization is considered isotropic, and a higher photoionization intensity leads to a more diffuse space charge distribution, resulting in an expanded head radius.

In contrast, for negative streamers, the head-expansion effect is slight and limited, as shown in Fig. \ref{fig.sp}(d-f) and \ref{fig.sp}(j-l). This is because, in negative streamers, electrons generated by impact ionization tend to drift toward the streamer front, naturally broadening the space charge region as well as the streamer head radius. As a result, the additional contribution from photoionization is less significant.

\noindent\hspace*{2em}\textbf{\textit{Head Field}} The \(50 \times S_{\text{ph}}\) case has a significant impact on the electric field strength at the positive streamer head. Specifically, the field strength decreases from approximately 1500 Td in the \(1 \times S_{\text{ph}}\) case (see Fig. \ref{fig.sp}(c)) to around 700 Td with \(50 \times S_{\text{ph}}\) (see Fig. \ref{fig.sp}(i)). This reduction is attributed to the enlarged streamer head radius resulting from stronger photoionization, which lowers the space charge density and reduces the related field distortion. 

In contrast, for negative streamers, the above-mentioned effect is limited. After 0.6 ns, \(50 \times S_{\text{ph}}\) case consistently enhances the electric field at the streamer head. 
This is because a higher photoionization intensity does not significantly affect the head radius (fundamentally the space charge distribution), but instead supplements the amount of space charge in the head, thereby enhancing the field distortion. However, due to the inherently large radius of the negative streamer, the contribution of the \(50 \times S_{\text{ph}}\) case to the increase in space charge density, is present but slight, leading to an increase of less than 30 Td.

\section{\label{4}discussion on the role of photoionization in streamer branching}

It should be noted that streamer branching is a complex phenomenon that may result from multiple mechanisms. These include local mechanisms—in essence Laplacian instability reported by Ebert and collaborators \cite{PhysRevLett.88.174502,PhysRevE.73.065401,Ebert_2006}; as well as nonlocal stochastic mechanisms, such as photoionization reported by Nijdam and collaborators \cite{Guo_2024,Dijcks_2023,Li_2020}. The deterministic 2D fluid model in this paper aims to investigate the general physics in highly non-uniform fields with low computational cost; however, this approach, as a rule, cannot capture the nonlocal stochastic process, as discussed in Refs. \cite{10.1063/5.0037669,Shao_2025} and references therein.

Therefore, based on insights from the literature, we provide a brief, though not exhaustive, discussion on the role of photoionization in streamer branching, focusing on its stochastic nature. In 1939, Raether \cite{H.Raether} first proposed a theory for streamer propagation using the cloud chamber experiment, then it was developed into a branching concept in extensive textbooks, e.g., Raizer \cite{FDLFSD}, Liang \textit{et al.} \cite{gaodianya} and Shao and Yan \cite{shao2015gas}, which reported that stochastic photoelectrons ahead of a positive streamer could initiate avalanches, then propagated back toward the streamer head and gave rise to new branches. More recent research \cite{Wormeester_2010} reported that rather weak photoionization led to a steep electron density drop at the front edge of the positive streamer head, such that the density should be treated probabilistically rather than as a continuous term. This analysis explained the stochastic, feather-like morphology of streamers observed in the experiment of \cite{Nijdam_2010}. Numerical research \cite{Bagheri_2019,Xiong_2014} showed that discretization of photoionization revealed the dependence of branching on photoionization intensity: lower intensity enhanced stochastic fluctuations and promoted branching. Experimental research \cite{4483864,PhysRevE.71.016407} further confirmed that increasing gas pressure—fundamentally reducing photoionization intensity—could increase branching frequency and randomness.

In summary, we would remind readers that, for quantitative predictions—particularly for estimating breakdown voltage in the context of insulating gases for high voltage technology, where multiple streamer events commonly occur \cite{10.1063/5.0223522,arxiv,Guo_2023,10.1063/5.0186055}—high fidelity, fully 3D modeling may provide higher accuracy \cite{Guo_2023home,TEUNISSEN2025109733,Wang_2023,Marskar_2025,Marskar_2024123}.

\begin{table*}[t]
  \centering\renewcommand{\arraystretch}{1.5}
  \caption{Summary of Helmholtz parameters (1 - 15 atm)}
  \label{tab:abc}
  \begin{tabular}{lcccccc}
    \toprule
    $p$ & $A_1 \, [\text{cm}^{-2} \cdot \text{Torr}^{-2}]$ & $A_2 \, [\text{cm}^{-2} \cdot \text{Torr}^{-2}]$ & $A_3 \, [\text{cm}^{-2} \cdot \text{Torr}^{-2}]$ & $\lambda_1 \, [\text{cm}^{-1} \cdot \text{Torr}^{-1}]$ & $\lambda_2 \, [\text{cm}^{-1} \cdot \text{Torr}^{-1}]$ & $\lambda_3 \, [\text{cm}^{-1} \cdot \text{Torr}^{-1}]$ \\
    \midrule\hline
    1 atm (760 Torr) & 123.1742 & 391.184 & 15.01558 & 3.27186 & 0.10233 & 1.00209 \\ \hline
    1.2 atm (912 Torr) & 176.449 & 575.522 & 23.0066 & 3.73336 & 0.08825 & 0.95739 \\\hline
    1.5 atm (1140 Torr) & 130.891 & 435.974 & 16.83386 & 3.38667 & 0.08717 & 0.94058 \\\hline
    1.8 atm (1368 Torr) & 136.5194 & 485.352 & 18.16002 & 3.64763 & 0.08816 & 0.94656 \\\hline
    2 atm (1520 Torr) & 17.3138 & 70.7354 & 137.1824 & 23.42125 & 0.09455 & 0.99277 \\\hline
    2.2 atm (1672 Torr) & 118.389 & 410.802 & 14.78442 & 3.22999 & 0.08188 & 0.95452 \\\hline
     2.5 atm (1900 Torr) & 14.23616 & 67.1082 & 117.1222 & 20.64586 & 0.09842 & 1.03453 \\\hline
    2.8 atm (2128 Torr) & 17.87222 & 74.4014 & 134.727 & 24.94662 & 0.09032 & 0.9635 \\\hline
    3 atm (2280 Torr) & 18.48346 & 75.1564 & 145.6424 & 26.46103 & 0.08932 & 0.98039 \\\hline
    4 atm (3040 Torr) & 16.05432 & 71.6138 & 125.5196 & 23.99987 & 0.08036 & 0.9819 \\\hline
    5 atm (3800 Torr) & 97.8708 & 376.28 & 9.89136 & 2.59363 & 0.0488 & 0.95394 \\\hline
    6 atm (4560 Torr) & 102.255 & 429.144 & 11.81758 & 3.01996 & 0.0514 & 0.89034 \\\hline
    7 atm (5320 Torr) & 11.31322 & 56.4932 & 110.1828 & 23.51018 & 0.04078 & 0.91384 \\\hline
     8 atm (6080 Torr) & 100.7306 & 471.792 & 12.0545 & 3.38219 & 0.06079 & 0.9569 \\\hline
    9 atm (6840 Torr) & 72.1056 & 366.746 & 9.2468 & 3.16824 & 0.062 & 0.99745 \\\hline
    10 atm (7600 Torr) & 11.15456 & 68.3094 & 92.7574 & 23.69784 & 0.05831 & 0.98973 \\\hline
    11 atm (8360 Torr) & 9.53146 & 63.206 & 77.7144 & 20.92194 & 0.04678 & 0.92143 \\\hline
    12 atm (9120 Torr) & 9.52426 & 65.7852 & 76.0054 & 21.50339 & 0.05005 & 0.95806 \\\hline
    13 atm (9880 Torr) & 8.81054 & 63.7478 & 71.1014 & 20.7923 & 0.0441 & 0.92392 \\\hline
    14 atm (10640 Torr) & 8.38072 & 67.2086 & 52.4864 & 17.27256 & 0.04868 & 0.93048 \\\hline
    15 atm (11400 Torr) & 12.48002 & 78.1596 & 115.5416 & 34.39515 & 0.04864 & 0.97245\\\hline
    \bottomrule
  \end{tabular}
\end{table*}

\section{\label{5}Concise Summary of Computational Photoionization Model and Helmholtz Parameters (1 - 15 atm)}

We adjust the pressure \(p\) during model development, and the same modeling procedure is extended to determine the Helmholtz parameters under high-pressure conditions. For direct reference, we provide a concise summary of the necessary descriptions of the SF$_6$ photoionization model and the three-term Helmholtz equation parameters within the pressure range of 1 - 15 atm as below.

The theoretical foundation of the SF\(_6\) photoionization model is based on Zheleznyak's classical photoionization model \cite{ZK} and Pancheshnyi's analytical model \cite{Pancheshnyi_2015}, while the numerical computation follows the Helmholtz equation model proposed by Luque \textit{et al.} \cite{10.1063/1.2435934} (and in parallel by Bourdon \textit{et al.} \cite{Bourdon_2007}). The photoionization source term is expressed:

\begin{equation}
   \label{Eqphoto99}
S_{\mathrm{ph}}(\mathbf{r})=\sum_{j} S_{\mathrm{ph}}^{j}(\mathbf{r}), j=1,2,3
\end{equation}

, where \(S_{\mathrm{ph}}^j(\mathbf{r}), j=1,2,3\) is obtained by solving the three-term Helmholtz equations:

\begin{equation}
   \label{Eqphoto101}
\nabla^2 S_{\mathrm{ph}}^j(\mathbf{r})-\left(\lambda_j p\right)^2 S_{\mathrm{ph}}^j(\mathbf{r})=-A_j p^2 I(\mathbf{r}), j=1,2,3
\end{equation}

, where \(I\)(r) is expressed:

\begin{equation}
   \label{Eqphoto102}
I(\mathbf{r})=\frac{p_{\mathrm{q}}}{p+p_{\mathrm{q}}} \xi \frac{v_{\mathrm{u}}}{v_{\mathrm{i}}} S_{\mathrm{i}}(\mathbf{r})
\end{equation}

, where $\frac{v_{\text{u}}}{v_i} \approx 0.00001$ for classical fluid model under the local field approximation, \(p_\text{q}\approx 2 \,\text{Torr}\), \(\xi\approx0.25\). \( S_{\mathrm{i}}(\boldsymbol{r}) = \alpha n_{\mathrm{e}} \left|\mu_{\mathrm{e}} \boldsymbol{E}\right| \) denotes the impact ionization rate. These parameters are considered pressure-independent, namely, they remain the same under different pressure conditions. In contrast, the fitted parameters of the three-term Helmholtz equation are pressure-dependent, with the corresponding values at different pressures listed in Table. \ref{tab:abc}.

For each pressure level, parameters are tested in plasma fluid simulations, ensuring spatial numerical convergence and the prediction of general coherent structures of single SF$_6$ streamer. However, only the case at \( p = 1 \, \text{atm} \) has been validated experimentally, whereas \( p > 1 \, \text{atm} \) parameters remain unvalidated. At high pressures, SF$_6$ DC breakdown is expected to involve multiple streamer channels, making 2D simulations unsuitable for direct comparison with experiment; 3D simulations would be required as already discussed in Section \ref{4}.

Nevertheless, we argue that the high-pressure parameters remain reasonable since theoretically, pressure $p$ affects only $\frac{g(R)}{p}$, which has already been incorporated into the modeling procedure, while $p_\text{q}$, $\frac{v_{\text{u}}}{v_i}$, and $\xi$ are unaffected. Interested readers could combine fully 3D simulations and experiments for validation; and, if necessary, further refine the photoionization model parameters such as \( \frac{v_{\text{u}}}{v_i} \), etc. Despite uncertainties, the high-pressure parameters in this paper can still serve as a useful reference.

\section{\label{6}Conclusions}

In this paper, we develop a computational model for SF\(_6\) photoionization and perform comparative studies. The key findings are summarized as follows:

(1) The theoretical foundation of the SF\(_6\) photoionization model is based on Zheleznyak's classical photoionization model \cite{ZK} and Pancheshnyi's analytical model \cite{Pancheshnyi_2015}, while the numerical computation follows the Helmholtz equation model proposed by Luque \textit{et al.} \cite{10.1063/1.2435934} (and in parallel by Bourdon \textit{et al.} \cite{Bourdon_2007}). The detailed theoretical modeling process, as well as comparison between experiment and simulation is provided and may serve as a reference for other gases in future studies. A concise summary of model parameters within the comprehensive pressure range of 1 - 15 atm is provided for direct reference.

(2) Comparative studies between the SF\(_6\) photoionization model proposed in this paper and commonly used simplified approaches demonstrate that the proposed model effectively captures the non-local effects of photoionization, leading to improved spatial numerical convergence and more accurate streamer structure predictions in SF\(_6\) streamer simulations.

(3) Comparative studies are conducted by modifying the multiplier of the photoionization source term in the continuity equation to assess the effects of photoionization intensity, comparing the results between \(50 \times S_{\text{ph}}\) and \(1 \times S_{\text{ph}}\). Regarding breakdown voltage prediction, the artificially increased photoionization intensity (\(50 \times S_{\text{ph}}\)) results in a significant underestimate of the positive streamer breakdown voltage, with errors exceeding 0.5 kV, while exerting only minor influence on negative streamers. Regarding streamer propagation dynamics, \(50 \times S_{\text{ph}}\) mitigates the shrinking at the positive streamer head and lowers the local field strength by more than 700 Td. In contrast, the negative streamer morphology remains fundamentally unaffected, and \(50 \times S_{\text{ph}}\) causes a slight enhancement in the local electric field (less than 30 Td).

\section{\label{7}outlook}

(1) The proposed photoionization model is limited to pure SF\(_6\) streamer. In the context of SF\(_6\) mixtures, the model may incorporate the specific properties of the buffer gas, including radiation, ionization, and other relevant processes. The absorption functions for each gas component may be calculated individually, and the corresponding proportions within the mixture may be considered. These extensions remain for further investigation in the future.

(2) It is important to note that other widely used computational models for photoionization, such as the three-group Eddington and SP3 approximations \cite{Ségur_2006}, the Green's function and propagator approach \cite{10.1063/1.3644953,Xiong_2010}, fast multipole method \cite{Lin_2020} and the explicit Monte Carlo approach \cite{MARSKAR2024112858,Marskar_2024}, may also offer effective predictions. Applying these methods to SF\(_6\) photoionization also remains for further investigation in the future.

(3) The model may involve uncertainty because it assumes a fixed shape of the VUV emission spectra, originally measured at an electron energy of 200 eV.

\section*{Acknowledgment}
The authors gratefully acknowledge funding support from the National Natural Science Foundation of China (Grant No. 52407176) and the Postdoctoral Fellowship Program of CPSF (Grant No. GZB20230326). The authors express sincere gratitude to the editors and referees for their constructive and insightful comments, which have substantially improved this manuscript. The authors thank all the professors at the Gas Discharge and Plasma Laboratory of Tsinghua University. Z. Feng thanks Dr. Peng Wang and Caomingzhe Si from the Department of Electrical Engineering, Tsinghua University, for fruitful discussions.

\section*{AUTHOR DECLARATIONS}
Conflict of Interest. The authors have no conflicts to disclose.
\section*{DATA AVAILABILITY}
All data that support the findings of this study are included within the article (and any supplementary files).

\section*{\label{A}Appendix A. Numerical Scheme}

\renewcommand{\thefigure}{A\arabic{figure}}
\setcounter{figure}{0}
\begin{figure}[b]
\centering
\includegraphics[width=8.5cm]{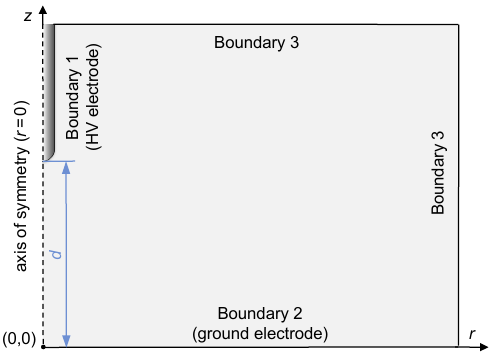}
\caption{\label{fig.geo} Geometry of the 2D axisymmetric simulation domain and the computational boundaries.}
\end{figure}

\noindent\hspace*{2em}\textbf{\textit{Geometry and Boundaries.}} As shown in Fig. \ref{fig.geo}, the high-voltage (HV) electrode is modeled as an elongated rod with a rounded tip. The rod electrode has a length of 5 mm, with its radius set to 0.1 mm in the main text and 0.5 mm in Appendix B. The curvature radius at the tip is equal to the radius of the rod. The gas gap distance between the rod tip and the ground electrode denotes \(d=1 \text{mm}\). The computational domain includes a sufficiently wide horizontal size to minimize edge effects. The boundary conditions are consistent with those specified in Ref. \cite{15}. The applied voltage \( U_0 \) is employed to Boundary 1.

\renewcommand{\thefigure}{B\arabic{figure}}
\setcounter{figure}{0}
\begin{figure*}[t]
\centering
\includegraphics[width=17cm]{Fig.B1.pdf}
\caption{\label{fig.B1} (a) Experimental system for capturing the discharge morphology, (b) 1:1 discharge electrode, and (c) ICCD photographs with an exposure time of 0.5 ns and the same optical gain, showing using pseudo-color.}
\end{figure*}

\noindent\hspace*{2em}\textbf{\textit{Governing Equations.}} The classical fluid model under the local field approximation is employed, considering only three charged species: electrons, positive ions, and negative ions. The number densities \(n_\text{e}\), \(n_\text{p}\), and \(n_\text{n}\) are calculated using continuity equations with the drift-diffusion approximation. The source terms of continuity equations include reaction source terms and photoionization source term. The reaction source terms include the contributions of impact ionization, attachment, detachment and recombination. The values of impact ionization coefficient, attachment coefficient, and electron transport coefficient are calculated from electron-neutral collision cross-sections by solving the 0D electron Boltzmann equation using the BOLSIG+ solver \cite{bosig+}. The electron-neutral collision reactions include elastic collisions, impact ionization generating \( \text{SF}_5^+ \), \( \text{SF}_4^+ \), \( \text{SF}_3^+ \), \( \text{SF}_2^+ \), \( \text{SF}^+ \), \( \text{S}^+ \), and \( \text{F}^+ \), attachment generating \( \text{SF}_6^- \), \( \text{SF}_5^- \), and \( \text{F}^- \), and 32 excitation reactions. The corresponding cross-sections are all taken from the Biagi database \cite{biagi2025}. The contribution of electron detachment in the source term is given by $(\frac{\delta}{N} ) \cdot N\cdot n_n \cdot\langle v_{\text{rel}} \rangle$, where $N$ is the number density of SF$_6$, $(\frac{\delta}{N})$ is the detachment coefficient taken from Figure 41 of Ref. \cite{10.1063/1.1288407}, and $\langle v_{\text{rel}} \rangle$ is the average relative velocity between SF$_6$ and the negative ions under the Maxwell-Boltzmann distribution. The corresponding velocity are $ 615 \, \text{m/s}$ for F$^-$, $295 \, \text{m/s}$ for \( \text{SF}_6^- \), and $306 \, \text{m/s}$ for \( \text{SF}_5^- \). In practical testing, the influence of different velocities on streamer properties and breakdown voltage is negligible, so we adopt $\langle v_{\text{rel}} \rangle = 615 \, \text{m/s}$. The value of recombination is taken from Ref. \cite{EN3602}. The values of ion transport coefficient are taken from Ref. \cite{EN3602}. The finite element method (FEM) is used, and the streamline diffusion technique is employed. Seed electrons are pre-ionized using a Gaussian distribution, consistent with Ref. \cite{arxiv}. In testing, different pre-ionization intensity and the electron detachment mechanism only affect streamer formation time, with negligible impact on propagation or breakdown voltage.

\noindent\hspace*{2em}\textbf{\textit{Grid Spacing.}}   
The adaptive refinement is employed, with the grid spacing refined automatically based on the error indicator:

\begin{equation*}
\sum_{i \in \{\text{e}, \text{p}, \text{n}\}} \sqrt{\left(\frac{\partial n_{i}}{\partial r}\right)^2 + \left(\frac{\partial n_{i}}{\partial z}\right)^2}
\end{equation*}
The minimum grid spacing during the refinement process is 0.1 \(\mu \text{m}\).

\noindent\hspace*{2em}\textbf{\textit{Time Step.}}   
The implicit backward differentiation formula (BDF) method is used, with a maximum BDF order of 2 and a minimum BDF order of 1. The relative tolerance is set to \(10^{-2}\). A parallel sparse direct solver (PARDISO) is selected as the direct linear system solver. Notably, the maximum time step is recommended 0.5 ps, and the rationale will be presented elsewhere.

\section*{\label{B}Appendix B. Comparison Between Experiment and Simulation to Estimate Parameter $\frac{v_\text{u}}{v_\text{i}}$}

The core of comparison between experiment and simulation is adjusting \( \frac{v_{\text{u}}}{v_{\mathrm{i}}} \) to match the positive breakdown voltage between simulations with experiments. The rationale for employing the positive streamer breakdown voltage for comparison is because photoionization intensity strongly affects it, as discussed in Section \ref{3.2}. 

\noindent\hspace*{2em}\textbf{\textit{1:1 Discharge Electrode.}} The rod-plate electrode creats a 1 mm SF\(_6\) gap under highly non-uniform field. The rod electrode is fabricated via CNC machining, in practice, for easier processing, the rod has a length of 5 mm, a radius of 0.5 mm, and a curvature radius of 0.5 mm at the tip. The tip geometry is verified using optical microscopy. The discharge electrode is positioned inside the vacuum chamber.

\begin{figure*}[t]
\centering
\includegraphics[width=17cm]{Fig.B2.pdf}
\caption{\label{fig.B2} (a) Experimental system for measuring the DC breakdown voltage, (b) the measured maximum breakdown voltage (max BD) and minimum breakdown voltage (min BD), as well as corresponding reasonable \( \frac{v_{\text{u}}}{v_{\mathrm{i}}} \) range, and (c) comparison between experiment and simulation using the positive breakdown voltage, as well as recommended \( \frac{v_{\text{u}}}{v_{\mathrm{i}}} \) value.}
\end{figure*}

\noindent\hspace*{2em}\textbf{\textit{ICCD Photograph of Discharge Morphology.}} Notably, to confirm the breakdown voltage equivalence between 2D simulation and DC experiment, the sufficient condition is that a single streamer crossing dominates the DC breakdown, without leader mechanisms or streamer branching. However, due to the statistical delay of DC discharge being much greater than nanoseconds, we are currently unable to synchronize the ICCD gate with the DC discharge event in the 1 mm gap, so we must acknowledge the equivalence cannot be directly verified in this paper. Here, within the scope of our current experimental conditions, we verify the below two necessary but insufficient conditions. The motivation is that if these two are satisfied, it would indicate that the possibility of a single streamer dominating the DC breakdown cannot be excluded, namely, a single streamer dominating the breakdown is also possible in the DC experiment. (1) Under nanosecond pulse voltage, breakdown of 1 mm 1 atm SF\(_6\) gap is dominated by a single streamer crossing. The experimental system for obtaining the discharge morphology is shown in Fig. \ref{fig.B1}(a). An ICCD camera equipped with a long-focus lens is used, with an exposure time of 0.5 ns. A pulse voltage with an amplitude of + 13 kV, a rise time of 500 ns, and with a repetitive pulse sequence with a frequency of 10 Hz, consisting of 100 pulses per sequence is applied to the rod electrode through a protective resistance. The breakdown moment, i.e., voltage drop moment, always occurs at the rising edge of the pulse voltage. The power source and the ICCD are synchronized via the DG645, with the ICCD gate set before the voltage drop moment. The chamber is filled with 1 atm of 99.99\% purity SF\(_6\). Voltage and current signals are measured using respective probes and recorded on the same oscilloscope. In practice, the first breakdown in each sequence always occurs after several pulses. We consider the first photograph which has discharge light signal in each sequence to represent the morphology of the pulse breakdown. Using the same and extremely high ICCD optical gain, the streamer morphology from different pulse sequences are shown in Fig. \ref{fig.B1}(c). The discharge morphology is characterized by a single streamer channel, different from the multi-channel morphology (see Ref. \cite{Omori_2024}). The varying streamer lengths lead to dispersion in the breakdown voltage, so a reasonable range should be provided when estimating \( \frac{v_{\text{u}}}{v_{\mathrm{i}}} \). (2) Optical imaging experiment in long-gap DC discharge showed that the length of the first SF\(_6\) positive streamer at 1 atm is greater than 1 mm (see Fig. 9 in Ref. \cite{Bujotzek_2015}).

\noindent\hspace*{2em}\textbf{\textit{Recommend $\frac{v_\text{u}}{v_\text{i}} \approx 0.00001$ Using Positive Breakdown Voltage.}}
We follow Seeger and Clemen’s approach \cite{Seeger_2014} for measuring DC breakdown voltage, which involves separately measuring the so-called maximum breakdown voltage (max BD) and minimum breakdown voltage (min BD). The max BD corresponds to scenario where breakdown occurs immediately (within a few microseconds) after voltage application with a 254 nm UV lamp reducing statistical time lag, so the max BD means the first streamer corona causing breakdown without the leader mechanism. The min BD corresponds to scenario where breakdown occurs within a significant time lag (30 seconds) without any UV lamp, so the min BD means possible leader mechanism and multiple streamer branching causing the breakdown. The experimental system for measuring the breakdown voltage is shown in Fig. \ref{fig.B2}(a), where the power supply is replaced with a positive polarity DC power source, and a 254 nm UV lamp is added, while the other conditions and discharge electrode remain unchanged. Before each measurement, the system is rested for 30 minutes. Only the first breakdown event is recorded, with fresh SF$_6$ gas, a new rod electrode, and a re-polished plane electrode for each measurement. A total of 8 sets are conducted, each with sperate measurement for max BD and min BD, respectively. The measurement results are shown in Fig. \ref{fig.B2}(b), and the min BD and max BD are almost identical. This further suggests that DC breakdown in a 1 mm, 1 atm SF$_6$ gap is likely dominated by a single streamer, without branching or leader mechanisms. We adopt the max BD (7.6 kV \(\sim\) 7.7 kV) as the benchmark for estimating $\frac{v_\text{u}}{v_\text{i}}$, as it corresponds to the breakdown caused by the first streamer corona without leader mechanism. In the simulation, we set a breakdown voltage error of 0.1 kV, yielding a reasonable $\frac{v_\text{u}}{v_\text{i}}$ range of $0.00001 \leq\frac{v_\text{u}}{v_\text{i}}\leq0.00004$, corresponding to the upper and lower limits of the max BD, as shown in Fig. \ref{fig.B2}(b). In this paper, we recommend employing $\frac{v_\text{u}}{v_\text{i}} \approx 0.00001$ (corresponding to +7.7 kV max BD, see Fig. \ref{fig.B2}(c)), because this value results in a positive streamer radius that aligns with experimental measurements (see Section. \ref{3.2.2}). In summary, the model parameter is recommended as $\frac{v_\text{u}}{v_\text{i}} \approx 0.00001$ for classical fluid model under the local field approximation.

\noindent\hspace*{2em}\textbf{\textit{Sources of Uncertainty.}}
There may be potential uncertainty in the experiment, such as electrode material, surface conditions, gas purity, ambient temperature, and pre-ionization intensity. Therefore, we must acknowledge that the quantitative estimation of $\frac{v_\text{u}}{v_\text{i}}$ is dependent on the specific experimental conditions. Also, there may be potential uncertainty in the model input data. Besides, for the local mean energy approximation (see Refs. \cite{Babaeva_20091,Levko_2017,Jiang_2023,Kong_2023} for detailed governing equations) or extended fluid model under the local field approximation (see Refs. \cite{Niknezhad_20212,Teunissen_2020,Soloviev_2009,Guo_2025} for detailed governing equations), the $\frac{v_\text{u}}{v_\text{i}}$ might vary, and if necessary, we will present detailed investigation elsewhere.

\Large
\bibliography{references}

@article{Li_2025,
doi = {10.1088/1361-6463/ae1043},
url = {https://doi.org/10.1088/1361-6463/ae1043},
year = {2025},
month = {oct},
publisher = {IOP Publishing},
volume = {58},
number = {44},
pages = {445501},
author = {Li, Xuan and Yang, Ning and Yin, Yichun and Wang, Yuan and Ren, Hanwen and Wang, Jian and Li, Qingmin},
title = {Simulation methodology for the dynamic adsorption effect of micro/nano-dust in gas-insulated switchgears and transmission lines and evaluation of multimode diffusion-induced explosions},
journal = {J. Phys. D: Appl. Phys.},
}

@article{Kong_2023,
doi = {10.1088/1361-6595/acfd5b},
url = {https://doi.org/10.1088/1361-6595/acfd5b},
year = {2023},
month = {oct},
publisher = {IOP Publishing},
volume = {32},
number = {10},
pages = {105004},
author = {Kong, Xianghao and Li, Sisi and Li, Haoyi and Yang, Weimin and Yang, Dezheng and Ning, Wenjun and Wang, Ruixue},
title = {Distribution patterns of reactive species in the interaction between atmospheric pressure plasma jet and fiber membrane},
journal = {Plasma Sources Sci. Technol.},
}

@article{Jiang_2023,
doi = {10.1088/1361-6463/acb3dc},
url = {https://doi.org/10.1088/1361-6463/acb3dc},
year = {2023},
month = {feb},
publisher = {IOP Publishing},
volume = {56},
number = {8},
pages = {085201},
author = {Jiang, Yuanyuan and Wang, Yanhui and Zhang, Jiao and Wang, Dezhen},
title = {Numerical study of helium atmospheric pressure plasma jet interacting with a wavy substrate surface with different dielectric constants and curvature radius},
journal = {J. Phys. D: Appl. Phys.},
}

@article{Levko_2017,
doi = {10.1088/1361-6595/aa5403},
url = {https://dx.doi.org/10.1088/1361-6595/aa5403},
year = {2017},
month = {feb},
publisher = {IOP Publishing},
volume = {26},
number = {3},
pages = {035003},
author = {Dmitry Levko and Laxminarayan L Raja},
title = {Fluid versus global model approach for the modeling of active species production by streamer discharge},
journal = {Plasma Sources Sci. Technol.},
}

@article{Guo_2025,
doi = {10.1088/1361-6463/ae1597},
url = {https://doi.org/10.1088/1361-6463/ae1597},
year = {2025},
month = {oct},
publisher = {IOP Publishing},
volume = {58},
number = {44},
pages = {445205},
author = {Guo, Yulin and Zhang, Yaqi and Zhu, Yifei and Jeanney, Pascal and Brisset, Alexandra and Sun, Anbang and Tardiveau, Pierre},
title = {Numerical investigation of atmospheric positive diffuse discharge mode transition: the effect of voltage rise time},
journal = {J. Phys. D: Appl. Phys.},
}

@article{Kourtzanidis_2025,
doi = {10.1088/1361-6463/adcace},
url = {https://doi.org/10.1088/1361-6463/adcace},
year = {2025},
month = {apr},
publisher = {IOP Publishing},
volume = {58},
number = {19},
pages = {195202},
author = {Kourtzanidis, Konstantinos and Starikovskaia, Svetlana M},
title = {Spatiotemporal dynamics of nanosecond pulsed discharge in the form of a fast ionization wave: self-consistent two-dimensional modeling and comparison with experiments under negative and positive polarity},
journal = {J. Phys. D: Appl. Phys.},
}

@article{Niknezhad_20212,
doi = {10.1088/1361-6595/ac24d2},
url = {https://doi.org/10.1088/1361-6595/ac24d2},
year = {2021},
month = {oct},
publisher = {IOP Publishing},
volume = {30},
number = {10},
pages = {105001},
author = {Niknezhad, M and Chanrion, O and Holbøll, J and Neubert, T},
title = {Dynamics of negative coronas in airflow},
journal = {Plasma Sources Sci. Technol.},
}

@article{Lin_2020,
doi = {10.1088/1361-6595/abc6f5},
url = {https://doi.org/10.1088/1361-6595/abc6f5},
year = {2020},
month = {dec},
publisher = {IOP Publishing},
volume = {29},
number = {12},
pages = {125010},
author = {Lin, Bo and Zhuang, Chijie and Cai, Zhenning and Zeng, Rong and Bao, Weizhu},
title = {Accurate and efficient calculation of photoionization in streamer discharges using fast multipole method},
journal = {Plasma Sources Sci. Technol.},
}

@article{Soloviev_2009,
doi = {10.1088/0022-3727/42/12/125208},
url = {https://doi.org/10.1088/0022-3727/42/12/125208},
year = {2009},
month = {jun},
publisher = {},
volume = {42},
number = {12},
pages = {125208},
author = {Soloviev, V R and Krivtsov, V M},
title = {Surface barrier discharge modelling for aerodynamic applications},
journal = {J. Phys. D: Appl. Phys.},
}

@article{Teunissen_2020,
doi = {10.1088/1361-6595/ab6757},
url = {https://doi.org/10.1088/1361-6595/ab6757},
year = {2020},
month = {jan},
publisher = {IOP Publishing},
volume = {29},
number = {1},
pages = {015010},
author = {Teunissen, Jannis},
title = {Improvements for drift-diffusion plasma fluid models with explicit time integration},
journal = {Plasma Sources Sci. Technol.},
}

@article{FPinnekamp_1983,
doi = {10.1088/0022-3727/16/7/019},
url = {https://doi.org/10.1088/0022-3727/16/7/019},
year = {1983},
month = {jul},
publisher = {},
volume = {16},
number = {7},
pages = {1293},
author = {F Pinnekamp and L Niemeyer},
title = {Qualitative model of breakdown in SF6 in inhomogeneous gaps},
journal = {J. Phys. D: Appl. Phys.},
}

@article{10.1063/5.0295511,
    author = {Shang, Wei and Su, Jiancang and Zeng, Bo and Cheng, Jie and Li, Rui and Xu, Xiudong and Yu, Binxiong and Gao, Mingzhu and Li, Yongdong},
    title = {All sealed MV, kA class self-breakdown multi-stage gas switch with low breakdown stability and long operation lifetime},
    journal = {Rev. Sci. Instrum.},
    volume = {96},
    number = {11},
    pages = {114702},
    year = {2025},
    month = {11},
    issn = {0034-6748},
    doi = {10.1063/5.0295511},
    url = {https://doi.org/10.1063/5.0295511},
}

@article{10.1063/5.0292765,
    author = {Naidis, G. V. and Babaeva, N. Yu.},
    title = {Positive streamers in air: Effect of the average applied field},
    journal = {Phys. Plasmas},
    volume = {32},
    number = {10},
    pages = {104501},
    year = {2025},
    month = {10},
    issn = {1070-664X},
    doi = {10.1063/5.0292765},
    url = {https://doi.org/10.1063/5.0292765},
}

@article{Starikovskiy_2022,
doi = {10.1088/1361-6595/aca04c},
url = {https://doi.org/10.1088/1361-6595/aca04c},
year = {2022},
month = {nov},
publisher = {IOP Publishing},
volume = {31},
number = {11},
pages = {114009},
author = {Starikovskiy, A Yu and Bazelyan, E M and Aleksandrov, N L},
title = {The influence of humidity on positive streamer propagation in long air gap},
journal = {Plasma Sources Sci. Technol.},
}

@article{Omori_2024,
doi = {10.1088/1361-6595/ad9cbe},
url = {https://dx.doi.org/10.1088/1361-6595/ad9cbe},
year = {2024},
month = {dec},
publisher = {IOP Publishing},
volume = {33},
number = {12},
pages = {125012},
author = {Omori, Kento and Ono, Ryo and Komuro, Atsushi},
title = {Streamer propagation in CO2 and N2/CO2 mixtures at atmospheric pressure},
journal = {Plasma Sources Sci. Technol.},
}

@article{BKDEA,
	author={Christoph Köhn and  Torsten Neubert and  Martin Füllekrug and  Ute Ebert and  Sander Nijdam and  Olivier Chanrion and Nikolai Østgaard and  Martino Marisaldi and  Serge Soula and  Joan Montanyà and  Francisco Gordillo-Vázquez and Alejandro Luque and  Jannis Teunissen and  Joachim Holbøll and  Alec Bennett and  Paul Smith and  Victor Lorenzo and Hugh J. Christian and  SørenF. Madsen and  Diana Mihailova and Jean-François Boissin and  Stéphane Pedeboy and  Laure Chaumat and Matthias Heumesser and Krystallia Dimitriadou and Carolina Maiorana and  Simon Ghilain and Zaida Gomez Kuri and  Adam Peverell and  Michele Urbani and  Thi Ny Kieu and  Andy Martinez and  Hani Francisco and  Mojtaba Niknezhad and  Miguel B. Teixeira-Gomes and  Andrea Pizzuti and  Marcelo Arcanjo and  Shahriar Mirpour and  Xue Bai and  Victor Reglero },
	title={Recent Results on Science and Innovation Related to Electrical Processes of Thunderstorms},
	journal={Surv. Geophys.},
doi = {10.1007/s10712-025-09891-x},
url={http://iopscience.iop.org/article/10.1088/1361-6463/add944},
	year={2025},
month = {August},
volume = {46},
pages = {753},
}

@article{10.1088/1361-6463/add944,
doi = {10.1088/1361-6463/add944},
url = {https://dx.doi.org/10.1088/1361-6463/add944},
year = {2025},
month = {may},
publisher = {IOP Publishing},
volume = {58},
number = {23},
pages = {233001},
author = {Zhuang, Weijian and Liang, Zuodong and Yi, Yong and Qin, Weiqi and Liang, Fangwei and Fan, Xianhao and Ma, Tiejun and Hu, Jun and Li, Chuanyang and Zhang, Bo and He, Jinliang},
title = {Metallic particles in DC gas-insulated transmission lines},
journal = {J. Phys. D: Appl. Phys.},
}

@article{Xiong_2010,
doi = {10.1088/0022-3727/43/50/505204},
url = {https://dx.doi.org/10.1088/0022-3727/43/50/505204},
year = {2010},
month = {dec},
publisher = {},
volume = {43},
number = {50},
pages = {505204},
author = {Xiong, Zhongmin and Kushner, Mark J},
title = {Surface corona-bar discharges for production of pre-ionizing UV light for pulsed high-pressure plasmas},
journal = {J. Phys. D: Appl. Phys.},
}

@article{Marskar_2024123,
doi = {10.1088/1361-6595/ad29c0},
url = {https://dx.doi.org/10.1088/1361-6595/ad29c0},
year = {2024},
month = {feb},
publisher = {IOP Publishing},
volume = {33},
number = {2},
pages = {025024},
author = {Marskar, R},
title = {Genesis of column sprites: formation mechanisms and optical structures},
journal = {Plasma Sources Sci. Technol.},
}

@article{Marskar_2025,
doi = {10.1088/1361-6463/adbe87},
url = {https://dx.doi.org/10.1088/1361-6463/adbe87},
year = {2025},
month = {mar},
publisher = {IOP Publishing},
volume = {58},
number = {18},
pages = {185201},
author = {Marskar, R},
title = {Towards quantitative partial discharge simulations},
journal = {J. Phys. D: Appl. Phys.},
}

@article{Wang_2023,
doi = {10.1088/1361-6595/ace9fa},
url = {https://dx.doi.org/10.1088/1361-6595/ace9fa},
year = {2023},
month = {aug},
publisher = {IOP Publishing},
volume = {32},
number = {8},
pages = {085007},
author = {Zhen Wang and Siebe Dijcks and Yihao Guo and Martijn van der Leegte and Anbang Sun and Ute Ebert and Sander Nijdam and Jannis Teunissen},
title = {Quantitative modeling of streamer discharge branching in air},
journal = {Plasma Sources Sci. Technol.},
}

@article{TEUNISSEN2025109733,
title = {Data-driven reduced modeling of streamer discharges in air},
journal = {Comput. Phys. Commun},
volume = {315},
pages = {109733},
year = {2025},
issn = {0010-4655},
doi = {https://doi.org/10.1016/j.cpc.2025.109733},
url = {https://www.sciencedirect.com/science/article/pii/S0010465525002358},
author = {Jannis Teunissen and Alejandro Malagón-Romero},
}

@article{Guo_2023home,
doi = {10.1088/1361-6595/acf87d},
url = {https://dx.doi.org/10.1088/1361-6595/acf87d},
year = {2023},
month = {sep},
publisher = {IOP Publishing},
volume = {32},
number = {9},
pages = {095015},
author = {Guo, Baohong and Ebert, Ute and Teunissen, Jannis},
title = {3D modeling of positive streamers in air with inhomogeneous density},
journal = {Plasma Sources Sci. Technol.},
}

@article{Guo_2023,
doi = {10.1088/1361-6595/ad0570},
url = {https://dx.doi.org/10.1088/1361-6595/ad0570},
year = {2023},
month = {nov},
publisher = {IOP Publishing},
volume = {32},
number = {11},
pages = {115001},
author = {Baohong Guo and Ute Ebert and Jannis Teunissen},
title = {3D particle-in-cell simulations of negative and positive streamers in C\(_4\)F\(_7\)N–CO\(_2\) mixtures},
journal = {Plasma Sources Sci. Technol.},
}

@ARTICLE{4483864,
  author={Nudnova, Maria M. and Starikovskii, Andrey Yu.},
  journal={IEEE Trans. Plasma Sci.}, 
  title={Development of streamer flash initiated by HV pulse with nanosecond rise time}, 
  year={2008},
  volume={36},
  number={4},
  pages={896-897},
  doi={10.1109/TPS.2008.920289}}

@article{Bagheri_2019,
doi = {10.1088/1361-6595/ab1331},
url = {https://dx.doi.org/10.1088/1361-6595/ab1331},
year = {2019},
month = {apr},
publisher = {IOP Publishing},
volume = {28},
number = {4},
pages = {045013},
author = {Bagheri, B and Teunissen, J},
title = {The effect of the stochasticity of photoionization on 3D streamer simulations},
journal = {Plasma Sources Sci. Technol.},
}

@article{Nijdam_2010,
doi = {10.1088/0022-3727/43/14/145204},
url = {https://dx.doi.org/10.1088/0022-3727/43/14/145204},
year = {2010},
month = {mar},
publisher = {},
volume = {43},
number = {14},
pages = {145204},
author = {Nijdam, S and van de Wetering, F M J H and Blanc, R and van Veldhuizen, E M and Ebert, U},
title = {Probing photo-ionization: experiments on positive streamers in pure gases and mixtures},
journal = {J. Phys. D: Appl. Phys.},
}

@article{Wormeester_2010,
doi = {10.1088/0022-3727/43/50/505201},
url = {https://dx.doi.org/10.1088/0022-3727/43/50/505201},
year = {2010},
month = {dec},
publisher = {},
volume = {43},
number = {50},
pages = {505201},
author = {Wormeester, G and Pancheshnyi, S and Luque, A and Nijdam, S and Ebert, U},
title = {Probing photo-ionization: simulations of positive streamers in varying N2 : O2-mixtures},
journal = {J. Phys. D: Appl. Phys.},
}

@book{gaodianya,
  author    = {Xidong Liang and Yuanxiang Zhou and Rong Zeng},
  title     = {High Voltage Engineering},
  publisher = {Tsinghua University Press},
  address   = {Beijing},
  year      = {2015},
  edition   = {1},
  note      = {(in Chinese)}

}

@book{shao2015gas,
  author    = {Tao Shao and Ping Yan},
  title     = {Atmospheric Gas Discharge and Plasma Applications},
  publisher = {Science Press},
  address   = {Beijing},
  year      = {2015},
  edition   = {1},
  note      = {(in Chinese)}
}

@book{FDLFSD,
  author    = {Y. P. Raizer},
  title     = {Gas Discharge Physics },
  edition   = {1},
  publisher = {Springer Press},
  year      = {1991},
  address   = {Berlin}
}

@article{H.Raether,
doi = {10.1007/BF01340229},
url = {https://doi.org/10.1007/BF01340229},
year = {1939},
month = {July },
publisher = {Springer},
volume = {112},
pages = {464},
author = {H. Raether},
title = {Die entwicklung der elektronenlawine in den funkenkanal: Nach beobachtungen in der nebelkammer},
journal = {Z. Phys.},
}

@article{Shao_2025,
doi = {10.1088/1361-6595/adfc0b},
url = {https://dx.doi.org/10.1088/1361-6595/adfc0b},
year = {2025},
month = {aug},
publisher = {IOP Publishing},
volume = {34},
number = {8},
pages = {085016},
author = {Shao, Xiao and Lacoste, Deanna A and Im, Hong G},
title = {A unified fluid model for nonthermal plasmas and reacting flows},
journal = {Plasma Sources Sci. Technol.},
}

@article{Li_2020,
doi = {10.1088/1361-6595/ab73de},
url = {https://dx.doi.org/10.1088/1361-6595/ab73de},
year = {2020},
month = {mar},
publisher = {IOP Publishing},
volume = {29},
number = {3},
pages = {03LT02},
author = {Li, Yuan and Dijcks, Siebe and Sun, Guangyu and Wen, Jiaye and Xu, Yaoyu and Zhang, Guanjun and Ebert, Ute and Nijdam, Sander},
title = {Characterizing streamer branching in N2–O2 mixtures by 2D peak-finding},
journal = {Plasma Sources Sci. Technol.},
}

@article{Dijcks_2023,
doi = {10.1088/1361-6595/acc821},
url = {https://dx.doi.org/10.1088/1361-6595/acc821},
year = {2023},
month = {apr},
publisher = {IOP Publishing},
volume = {32},
number = {4},
pages = {045004},
author = {Dijcks, Siebe and der Leegte, Martijn van and Nijdam, Sander},
title = {Imaging and reconstruction of positive streamer discharge tree structures},
journal = {Plasma Sources Sci. Technol.},
}

@article{Guo_2024,
doi = {10.1088/1361-6595/ad37bf},
url = {https://dx.doi.org/10.1088/1361-6595/ad37bf},
year = {2024},
month = {apr},
publisher = {IOP Publishing},
volume = {33},
number = {4},
pages = {045006},
author = {Yihao Guo and Sander Nijdam},
title = {Statistical analysis on branching characteristics of positive streamer discharges in N\(_2\)–O\(_2\) mixtures},
journal = {Plasma Sources Sci. Technol.},
}

@article{PhysRevE.73.065401,
  title = {Numerical convergence of the branching time of negative streamers},
  author = {Montijn, Carolynne and Ebert, Ute and Hundsdorfer, Willem},
  journal = {Phys. Rev. E},
  volume = {73},
  issue = {6},
  pages = {065401},
  numpages = {4},
  year = {2006},
  month = {Jun},
  publisher = {American Physical Society},
  doi = {10.1103/PhysRevE.73.065401},
  url = {https://link.aps.org/doi/10.1103/PhysRevE.73.065401}
}

@article{PhysRevLett.88.174502,
  title = {Spontaneous Branching of Anode-Directed Streamers between Planar Electrodes},
  author = {Array\'as, Manuel and Ebert, Ute and Hundsdorfer, Willem},
  journal = {Phys. Rev. Lett.},
  volume = {88},
  issue = {17},
  pages = {174502},
  numpages = {4},
  year = {2002},
  month = {Apr},
  publisher = {American Physical Society},
  doi = {10.1103/PhysRevLett.88.174502},
  url = {https://link.aps.org/doi/10.1103/PhysRevLett.88.174502}
}

@article{Ebert_2006,
doi = {10.1088/0963-0252/15/2/S14},
url = {https://dx.doi.org/10.1088/0963-0252/15/2/S14},
year = {2006},
month = {apr},
publisher = {},
volume = {15},
number = {2},
pages = {S118},
author = {Ebert, U and Montijn, C and Briels, T M P and Hundsdorfer, W and Meulenbroek, B and Rocco, A and van Veldhuizen, E M},
title = {The multiscale nature of streamers},
journal = {Plasma Sources Sci. Technol.},
}

@article{CELIK2012566,
title = {Transition probabilities and radiative lifetimes of levels in F I},
journal = {At. Data Nucl. Data Tables},
volume = {98},
number = {4},
pages = {566-588},
year = {2012},
issn = {0092-640X},
doi = {https://doi.org/10.1016/j.adt.2012.01.003},
url = {https://www.sciencedirect.com/science/article/pii/S0092640X1200023X},
author = {Gültekin Çelik and Duygu Doğan and Şule Ateş and Mehmet Taşer},
}

@book{Ju,
  author    = {Yiguang Ju and Andrey Starikovskiy},
  title     = {Plasma Assisted Combustion and Chemical Processing},
  edition   = {1},
  publisher = {CRC Press},
  year      = {2025},
  address   = {Boca Raton}
}

@article{Li_2024,
doi = {10.1088/1361-6595/ad7d35},
url = {https://dx.doi.org/10.1088/1361-6595/ad7d35},
year = {2024},
month = {sep},
publisher = {IOP Publishing},
volume = {33},
number = {9},
pages = {095009},
author = {Li, Xiaoran and Dijcks, Siebe and Sun, Anbang and Nijdam, Sander and Teunissen, Jannis},
title = {Investigation of positive streamers in CO2: experiments and 3D particle-in-cell simulations},
journal = {Plasma Sources Sci. Technol.},
}

@misc{BSR,
  howpublished = {BSR database, https://www.lxcat.net},
  note = {Retrieved on August 22, 2025}
}

@article{arxiv,
  title = {${\mathrm{SF}}_{6}$ streamer breakdown induced by floating linear metal particles: Following streamers and side streamers},
  author = {Feng, Zihao and Zhang, Liyang and Wang, Xinxin and Zou, Xiaobing and Luo, Haiyun and Fu, Yangyang},
  journal = {Phys. Rev. Appl.},
  volume = {23},
  issue = {6},
  pages = {064039},
  numpages = {21},
  year = {2025},
  month = {Jun},
  publisher = {American Physical Society},
  doi = {10.1103/xslk-zb7d},
  url = {https://link.aps.org/doi/10.1103/xslk-zb7d}
}

@article{Bouwman_2025,
doi = {10.1088/1361-6595/adaf53},
url = {https://dx.doi.org/10.1088/1361-6595/adaf53},
year = {2025},
month = {feb},
publisher = {IOP Publishing},
volume = {34},
number = {2},
pages = {025015},
author = {Bouwman, Dennis and Teunissen, Jannis and Ebert, Ute},
title = {Macroscopic parameterization of positive streamer heads in air},
journal = {Plasma Sources Sci. Technol.},
}

@article{PhysRevE.71.016407,
  title = {Development of a cathode-directed streamer discharge in air at different pressures: Experiment and comparison with direct numerical simulation},
  author = {Pancheshnyi, S. and Nudnova, M. and Starikovskii, A.},
  journal = {Phys. Rev. E},
  volume = {71},
  issue = {1},
  pages = {016407},
  numpages = {12},
  year = {2005},
  month = {Jan},
  publisher = {American Physical Society},
  doi = {10.1103/PhysRevE.71.016407},
}

@article{SVPancheshnyi_2001,
doi = {10.1088/0022-3727/34/1/317},
url = {https://dx.doi.org/10.1088/0022-3727/34/1/317},
year = {2001},
month = {jan},
publisher = {},
volume = {34},
number = {1},
pages = {105},
author = {S V Pancheshnyi and S M Starikovskaia and A Yu Starikovskii},
title = {Role of photoionization processes in propagation of cathode-directed streamer},
journal = {J. Phys. D: Appl. Phys.},
}

@article{Nudnova_2008,
doi = {10.1088/0022-3727/41/23/234003},
url = {https://dx.doi.org/10.1088/0022-3727/41/23/234003},
year = {2008},
month = {nov},
publisher = {},
volume = {41},
number = {23},
pages = {234003},
author = {Nudnova, M M and Starikovskii, A Yu},
title = {Streamer head structure: role of ionization and photoionization},
journal = {J. Phys. D: Appl. Phys.},
}

@article{SVPancheshnyi_2004,
doi = {10.1088/0963-0252/13/3/B01},
url = {https://dx.doi.org/10.1088/0963-0252/13/3/B01},
year = {2004},
month = {jul},
publisher = {},
volume = {13},
number = {3},
pages = {B1},
author = {S V Pancheshnyi and A Yu Starikovskii},
title = {Stagnation dynamics of a cathode-directed streamer discharge in air},
journal = {Plasma Sources Sci. Technol.},
}

@article{10.1063/5.0037669,
    author = {Starikovskiy, A. Yu. and Aleksandrov, N. L. and Shneider, M. N.},
    title = {Simulation of decelerating streamers in inhomogeneous atmosphere with implications for runaway electron generation},
    journal = {J. Appl. Phys.},
    volume = {129},
    number = {6},
    pages = {063301},
    year = {2021},
    month = {02},
    issn = {0021-8979},
    doi = {10.1063/5.0037669},
    url = {https://doi.org/10.1063/5.0037669},
}

@article{Xiong_2014,
doi = {10.1088/0963-0252/23/6/065041},
url = {https://dx.doi.org/10.1088/0963-0252/23/6/065041},
year = {2014},
month = {oct},
publisher = {IOP Publishing},
volume = {23},
number = {6},
pages = {065041},
author = {Xiong, Zhongmin and Kushner, Mark J},
title = {Branching and path-deviation of positive streamers resulting from statistical photon transport},
journal = {Plasma Sources Sci. Technol.},
}

@article{Babaeva_20091,
doi = {10.1088/0963-0252/18/3/035009},
url = {https://dx.doi.org/10.1088/0963-0252/18/3/035009},
year = {2009},
month = {may},
publisher = {},
volume = {18},
number = {3},
pages = {035009},
author = {Natalia Yu Babaeva and Mark J Kushner},
title = {Effect of inhomogeneities on streamer propagation: I. Intersection with isolated bubbles and particles},
journal = {Plasma Sources Sci. Technol.},
}

@article{10.1063/1.3644953,
    author = {Xiong, Zhongmin and Kushner, Mark J.},
    title = {Photo-triggering and secondary electron produced ionization in electric discharge ArF* excimer lasers},
    journal = {J. Appl. Phys.},
    volume = {110},
    number = {8},
    pages = {083304},
    year = {2011},
    month = {10},
    issn = {0021-8979},
    doi = {10.1063/1.3644953},
    url = {https://doi.org/10.1063/1.3644953},
}

@article{MARSKAR2024112858,
title = {Stochastic and self-consistent 3D modeling of streamer discharge trees with Kinetic Monte Carlo},
journal = {J. Comput. Phys.},
volume = {504},
pages = {112858},
year = {2024},
issn = {0021-9991},
doi = {https://doi.org/10.1016/j.jcp.2024.112858},
url = {https://www.sciencedirect.com/science/article/pii/S0021999124001074},
author = {Robert Marskar},
}

@article{Marskar_2024,
doi = {10.1088/1361-6595/ad28cf},
url = {https://dx.doi.org/10.1088/1361-6595/ad28cf},
year = {2024},
month = {feb},
publisher = {IOP Publishing},
volume = {33},
number = {2},
pages = {025023},
author = {R Marskar},
title = {A 3D kinetic Monte Carlo study of streamer discharges in \mathrm{CO_2}},
journal = {Plasma Sources Sci. Technol.},
}

@article{Ségur_2006,
doi = {10.1088/0963-0252/15/4/009},
url = {https://dx.doi.org/10.1088/0963-0252/15/4/009},
year = {2006},
month = {jul},
publisher = {},
volume = {15},
number = {4},
pages = {648},
author = {Ségur, P and Bourdon, A and Marode, E and Bessieres, D and Paillol, J H},
title = {The use of an improved Eddington approximation to facilitate the calculation of photoionization in streamer discharges},
journal = {Plasma Sources Sci. Technol.},
}

@article{Li_2022,
doi = {10.1088/1361-6595/ac7747},
url = {https://dx.doi.org/10.1088/1361-6595/ac7747},
year = {2022},
month = {jul},
publisher = {IOP Publishing},
volume = {31},
number = {6},
pages = {065011},
author = {Li, Xiaoran and Guo, Baohong and Sun, Anbang and Ebert, Ute and Teunissen, Jannis},
title = {A computational study of steady and stagnating positive streamers in N2–O2 mixtures},
journal = {Plasma Sources Sci. Technol.},
}

@article{Niknezhad_2021,
doi = {10.1088/1361-6595/ac3214},
url = {https://dx.doi.org/10.1088/1361-6595/ac3214},
year = {2021},
month = {nov},
publisher = {IOP Publishing},
volume = {30},
number = {11},
pages = {115014},
author = {Niknezhad, M and Chanrion, O and Holbøll, J and Neubert, T},
title = {Underlying mechanism of the stagnation of positive streamers},
journal = {Plasma Sources Sci. Technol.},
}

@article{Guo_2022123,
doi = {10.1088/1361-6595/ac8e2e},
url = {https://dx.doi.org/10.1088/1361-6595/ac8e2e},
year = {2022},
month = {sep},
publisher = {IOP Publishing},
volume = {31},
number = {9},
pages = {095011},
author = {Guo, Baohong and Li, Xiaoran and Ebert, Ute and Teunissen, Jannis},
title = {A computational study of accelerating, steady and fading negative streamers in ambient air},
journal = {Plasma Sources Sci. Technol.},
}

@Article{j5020018,
AUTHOR = {Boakye-Mensah, Francis and Bonifaci, Nelly and Hanna, Rachelle and Niyonzima, Innocent and Timoshkin, Igor},
TITLE = {Modelling of Positive Streamers in SF6 Gas under Non-Uniform Electric Field Conditions: Effect of Electronegativity on Streamer Discharges},
JOURNAL = {J},
VOLUME = {5},
YEAR = {2022},
NUMBER = {2},
PAGES = {255--276},
URL = {https://www.mdpi.com/2571-8800/5/2/18},
ISSN = {2571-8800},
DOI = {10.3390/j5020018}
}

@article{Seeger_2008123,
doi = {10.1088/0022-3727/41/18/185204},
url = {https://dx.doi.org/10.1088/0022-3727/41/18/185204},
year = {2008},
month = {aug},
publisher = {},
volume = {41},
number = {18},
pages = {185204},
author = {Seeger, M and Niemeyer, L and Bujotzek, M},
title = {Partial discharges and breakdown at protrusions in uniform background fields in SF6},
journal = {J. Phys. D: Appl. Phys.},
}

@article{10.1063/1.3629989,
    author = {Rose, D. V. and Welch, D. R. and Clark, R. E. and Thoma, C. and Zimmerman, W. R. and Bruner, N. and Rambo, P. K. and Atherton, B. W.},
    title = {Towards a fully kinetic 3D electromagnetic particle-in-cell model of streamer formation and dynamics in high-pressure electronegative gases},
    journal = {Physics of Plasmas},
    volume = {18},
    number = {9},
    pages = {093501},
    year = {2011},
    month = {09},
    issn = {1070-664X},
    doi = {10.1063/1.3629989},
    url = {https://doi.org/10.1063/1.3629989},
}

@ARTICLE{10969826,
  author={Zhong, Lipeng and Liu, Zulong and Du, Junxian and Yi, Shuang and Liang, Kaibin and Tang, Nian and Sun, Dongwei and Chen, She and Sun, Qiuqin and Wang, Feng},
  journal={IEEE Trans. Dielectr. Electr. Insul.}, 
  title={Investigation of Surface Flashover Mechanisms in SF6/N2 Mixtures under Nanosecond Pulsed}, 
  year={2025},
  volume={},
  number={},
  pages={1-1},
  doi={10.1109/TDEI.2025.3562163}}

@article{Wang_2025,
doi = {10.1088/1361-6463/ad9590},
url = {https://dx.doi.org/10.1088/1361-6463/ad9590},
year = {2024},
month = {nov},
publisher = {IOP Publishing},
volume = {58},
number = {6},
pages = {065203},
author = {Wang, Wei and Wang, Xinyan and Yang, Xin and Wu, Zhenyu and Xu, Hao and Meng, Yongpeng and Lv, Zepeng and Wu, Kai},
title = {Experimental and numerical study on surface streamer evolution in SF6/N2 gas mixtures},
journal = {J. Phys. D: Appl. Phys.},
}

@ARTICLE{10092907,
  author={Zhang, Zhaoqi and Song, Hui and Luo, Lingen and Sheng, Gehao and Jiang, Xiuchen},
  journal={IEEE Trans. Dielectr. Electr. Insul.}, 
  title={Comprehensive Simulation Research on Microprocess and Macroscopic Electromagnetic Signals of Surface Discharge in SF6}, 
  year={2023},
  volume={30},
  number={4},
  pages={1769-1778},
  doi={10.1109/TDEI.2023.3264960}}

@ARTICLE{10230229,
  author={Abbas, Muhammad Farasat and He, Yan Liang and Sun, Guang Yu and Sun, An Bang and Eldin, Elsayed Tag and Ghoneim, Sherif S. M.},
  journal={IEEE Access}, 
  title={Positive Streamer Initiation in SF6/CO2 Based on Zener’s Field Ionization}, 
  year={2023},
  volume={11},
  number={},
  pages={91767-91776},
  doi={10.1109/ACCESS.2023.3308688}}

@article{10.1063/5.0076343,
    author = {Zhang, Runming and Wang, Lijun and Liu, Jie and Lian, Zhuoxi},
    title = {Numerical simulation of breakdown properties and streamer development processes in SF6/CO2 mixed gas},
    journal = {AIP Adv.},
    volume = {12},
    number = {1},
    pages = {015003},
    year = {2022},
    month = {01},
    issn = {2158-3226},
    doi = {10.1063/5.0076343},
    url = {https://doi.org/10.1063/5.0076343},
}

@article{NYuBabaeva_2002,
doi = {10.1088/0022-3727/35/2/305},
url = {https://dx.doi.org/10.1088/0022-3727/35/2/305},
year = {2001},
month = {dec},
publisher = {},
volume = {35},
number = {2},
pages = {132},
author = {N Yu Babaeva and G V Naidis},
title = {Simulation of stepped propagation of positive
 streamers in SF6},
journal = {J. Phys. D: Appl. Phys.},
}

@ARTICLE{10400497,
  author={Sun, Yuanji and Li, Zhen and Liu, Ji and Gao, He and Zhang, Longfei and Wang, Shouming and Sun, Lei and Liu, Menglin and Li, Shengtao},
  journal={IEEE Trans. Dielectr. Electr. Insul.}, 
  title={Effects of Electric Field Nonuniformity on DC Positive Streamer Discharge of SF6/N2 Mixed Gas}, 
  year={2024},
  volume={31},
  number={4},
  pages={2019-2028},
  doi={10.1109/TDEI.2024.3354672}}

@ARTICLE{9099705,
  author={Luo, Bin and He, Hengxin and Cheng, Chen and Xia, Shengguo and Du, Weijie and Bian, Kai and Chen, Weijiang},
  journal={IEEE Trans. Dielectr. Electr. Insul.}, 
  title={Numerical simulation of the positive streamer propagation and chemical reactions in SF6/N2 mixtures under non-uniform field}, 
  year={2020},
  volume={27},
  number={3},
  pages={782-790},
  doi={10.1109/TDEI.2019.008551}}

@article{https://doi.org/10.1002/tee.24244,
author = {Seeger, Martin and Macedo, Felipe and Riechert, Uwe and Bujotzek, Markus and Hassanpoor, Arman and Häfner, Jürgen},
title = {Trends in High Voltage Switchgear Research and Technology},
journal = {IEEJ Trans. Electr. Electron. Eng.},
volume = {20},
number = {3},
pages = {322-338},
doi = {https://doi.org/10.1002/tee.24244},
url = {https://onlinelibrary.wiley.com/doi/abs/10.1002/tee.24244},
year = {2025}
}

@ARTICLE{90289,
  author={Niemeyer, L. and Ullrich, L. and Wiegart, N.},
  journal={IEEE Trans. Electr. Insul.}, 
  title={The mechanism of leader breakdown in electronegative gases}, 
  year={1989},
  volume={24},
  number={2},
  pages={309-324},
  doi={10.1109/14.90289}}

@article{doi:10.1139/p86-048,
author = {Forand, J. L. and Becker, K. and McConkey, J. W.},
title = {Dissociative excitation of SF6 by controlled electron impact},
journal = {Can. J. Phys.},
volume = {64},
number = {3},
pages = {269-276},
year = {1986},
doi = {10.1139/p86-048},
URL = {    
        https://doi.org/10.1139/p86-048}
}

@article{DMPHolland_1992,
doi = {10.1088/0953-4075/25/22/017},
url = {https://dx.doi.org/10.1088/0953-4075/25/22/017},
year = {1992},
month = {nov},
publisher = {},
volume = {25},
number = {22},
pages = {4823},
author = {D M P Holland and D A Shaw and A Hopkirk and M A MacDonald and S M McSweeney},
title = {A study of the absolute photoabsorption cross section and the photoionization quantum efficiency of sulphur hexafluoride from the ionization threshold to 420 A},
journal = {J. Phys. B: At. Mol. Opt. Phys.},
}

@misc{biagi2025,
  howpublished = {Biagi database, https://www.lxcat.net},
  note = {Retrieved on March 1, 2025}
}

@article{bosig+,
doi = {10.1088/0963-0252/14/4/011},
year = {2005},
month = {oct},
publisher = {},
volume = {14},
number = {4},
pages = {722},
author = {G J M Hagelaar and L C Pitchford},
title = {Solving the Boltzmann equation to obtain electron transport coefficients and rate coefficients for fluid models},
journal = {Plasma Sources Sci. Technol.},
}

@article{LRicheboeuf_1998,
doi = {10.1088/0022-3727/31/4/007},
url = {https://dx.doi.org/10.1088/0022-3727/31/4/007},
year = {1998},
month = {feb},
publisher = {},
volume = {31},
number = {4},
pages = {373},
author = {L Richeboeuf and S Pasquiers and M Legentil and V Puech},
title = {The influence of  and  molecules on discharge equilibrium and F-atom production in a phototriggered HF laser using},
journal = {J. Phys. D: Appl. Phys.},
}

@article{Pancheshnyi_2015,
doi = {10.1088/0963-0252/24/1/015023},
url = {https://dx.doi.org/10.1088/0963-0252/24/1/015023},
year = {2014},
month = {dec},
publisher = {IOP Publishing},
volume = {24},
number = {1},
pages = {015023},
author = {Pancheshnyi, Sergey},
title = {Photoionization produced by low-current discharges in O2, air, N2 and CO2},
journal = {Plasma Sources Sci. Technol.},
}

@article{Naidis_2006,
doi = {10.1088/0963-0252/15/2/010},
url = {https://dx.doi.org/10.1088/0963-0252/15/2/010},
year = {2006},
month = {mar},
publisher = {},
volume = {15},
number = {2},
pages = {253},
author = {Naidis, G V},
title = {On photoionization produced by discharges in air},
journal = {Plasma Sources Sci. Technol.},
}

@article{10.1063/1.465149,
    author = {Ying, J. F. and Daniels, T. A. and Mathers, C. P. and Zhu, H. and Leung, K. T.},
    title = {Absolute transition probability measurement of nondipole valence‐shell (7–70 eV) electronic transitions of SF6 by angle‐resolved electron energy loss spectroscopy},
    journal = {J. Chem. Phys.},
    volume = {99},
    number = {5},
    pages = {3390-3399},
    year = {1993},
    month = {09},
    issn = {0021-9606},
    doi = {10.1063/1.465149},
    url = {https://doi.org/10.1063/1.465149},
}

@article{zhu2020,
  author = {Zhu, Y. and Wu, Y. and Li, J.},
  title = {Photopic: Calculate photo-ionization functions and model coefficients for gas discharge simulations},
  year = {2020},
  journal = {arXiv:2005.10021},
}

@article{Ma_2024,
doi = {10.1088/1361-6595/ad5df6},
url = {https://dx.doi.org/10.1088/1361-6595/ad5df6},
year = {2024},
month = {jul},
publisher = {IOP Publishing},
volume = {33},
number = {7},
pages = {075012},
author = {Ma, Xiaochi and Bai, Luying and Zhu, Yifei and Jiang, Xinxian and Wu, Yun},
title = {Numerical investigation of discharge evolution and breakdown characteristics of ArF excimer lasers},
journal = {Plasma Sources Sci. Technol.},
}

@article{He_2024,
doi = {10.1088/1361-6595/ad6501},
url = {https://dx.doi.org/10.1088/1361-6595/ad6501},
year = {2024},
month = {aug},
publisher = {IOP Publishing},
volume = {33},
number = {8},
pages = {085008},
author = {He, Hengxin and Zhang, Wanxia and Liu, Lipeng and Luo, Bin and Chen, Ying and Zhang, Shiming and Xiao, Mian and Huang, Yubin and Chen, Shen},
title = {Effects of trace oxygen on the self-oscillation of positive glow corona in nitrogen near atmospheric pressure},
journal = {Plasma Sources Sci. Technol.},
}

@ARTICLE{10007898,
  author={Li, Chuanyang and Zhang, Changhong and Lv, Jinzhuang and Liang, Fangwei and Liang, Zuodong and Fan, Xianhao and Riechert, Uwe and Li, Zhen and Liu, Peng and Xue, Jianyi and Pan, Cheng and Chen, Geng and Zhang, Lei and Wang, Zheming and Lu, Wu and Liang, Hucheng and Pan, Zijun and Zhuang, Weijian and Mazzanti, Giovanni and Fabiani, Davide and Liu, Bo and Cao, Shaohua and Zhong, Jianying and Deng, Yuan and Nan, Zhenle and Tang, Jingen and He, Jinliang},
  journal={iEnergy}, 
  title={China's 10-year progress in DC gas-insulated equipment: From basic research to industry perspective}, 
  year={2022},
  volume={1},
  number={4},
  pages={400-433},
  doi={10.23919/IEN.2022.0050}}

@ARTICLE{10089517,
  author={Hallas, Martin and Neumann, Claus and Hinrichsen, Volker and Gross, Detlev},
  journal={IEEE Trans. Power Deliv.}, 
  title={Commissioning and Service Experience With a ±550 kV DC GIL Conducted in Frame of a CIGRE Prototype Installation Test}, 
  year={2023},
  volume={38},
  number={4},
  pages={2843-2853},
  doi={10.1109/TPWRD.2023.3263722}}

@article{Waters_2019,
doi = {10.1088/1361-6463/aae815},
url = {https://dx.doi.org/10.1088/1361-6463/aae815},
year = {2018},
month = {nov},
publisher = {IOP Publishing},
volume = {52},
number = {2},
pages = {025203},
author = {R T Waters},
title = {Electrical breakdown at high pressures: a Paschen law function and compressible gas dynamics},
journal = {J. Phys. D: Appl. Phys.}
}

@article{Seeger_2014,
doi = {10.1088/0022-3727/47/2/025202},
url = {https://dx.doi.org/10.1088/0022-3727/47/2/025202},
year = {2013},
month = {dec},
publisher = {IOP Publishing},
volume = {47},
number = {2},
pages = {025202},
author = {M Seeger and M Clemen},
title = {Partial discharges and breakdown in SF6 in the pressure range 25–150 kPa in non-uniform background fields},
journal = {J. Phys. D: Appl. Phys.}
}

@article{Seeger_2009,
doi = {10.1088/0022-3727/42/18/185205},
url = {https://dx.doi.org/10.1088/0022-3727/42/18/185205},
year = {2009},
month = {sep},
publisher = {},
volume = {42},
number = {18},
pages = {185205},
author = {M Seeger and L Niemeyer and M Bujotzek},
title = {Leader propagation in uniform background fields in SF6},
journal = {J. Phys. D: Appl. Phys.}
}

@article{Wu_2021,
doi = {10.1088/1361-6595/abd2cf},
url = {https://dx.doi.org/10.1088/1361-6595/abd2cf},
year = {2021},
month = {jan},
publisher = {IOP Publishing},
volume = {30},
number = {1},
pages = {015009},
author = {Zhicheng Wu and Binqi Lin and Xing Fan and Qiaogen Zhang and Licheng Li},
title = {Electric field dependence of SF6 nonlinear discharge characteristics: N-curve estimations},
journal = {Plasma Sources Sci. Technol.}
}

@ARTICLE{4074858,
  author={Cookson, Alan H. and Farish, Owen and L. Sommerman, George M.},
  journal={IEEE Trans. Power App. Syst.}, 
  title={Effect of Conducting Particles on AC Corona and Breakdown in Compressed SF6}, 
  year={1972},
  volume={PAS-91},
  number={4},
  pages={1329-1338},
  doi={10.1109/TPAS.1972.293262}}

@article{Wu_2019,
doi = {10.1088/1361-6595/ab32f4},
url = {https://dx.doi.org/10.1088/1361-6595/ab32f4},
year = {2019},
month = {aug},
publisher = {IOP Publishing},
volume = {28},
number = {8},
pages = {085018},
author = {Zhicheng Wu and Qiaogen Zhang and Lisong Zhang and Can Guo and Qiandong Du and Lei Pang},
title = {Neglected culprit of nonlinear discharge characteristics in SF6: shielding effect induced by positive glow corona discharge},
journal = {Plasma Sources Sci. Technol.},
}

@article{Nijdam_2020,
doi = {10.1088/1361-6595/abaa05},
url = {https://dx.doi.org/10.1088/1361-6595/abaa05},
year = {2020},
month = {oct},
publisher = {IOP Publishing},
volume = {29},
number = {10},
pages = {103001},
author = {Sander Nijdam and Jannis Teunissen and Ute Ebert},
title = {The physics of streamer discharge phenomena},
journal = {Plasma Sources Sci. Technol.}
}

@article{Francisco_2021,
doi = {10.1088/1361-6595/abdaa3},
url = {https://dx.doi.org/10.1088/1361-6595/abdaa3},
year = {2021},
month = {feb},
publisher = {IOP Publishing},
volume = {30},
number = {2},
pages = {025006},
author = {Hani Francisco and Behnaz Bagheri and Ute Ebert},
title = {Electrically isolated propagating streamer heads formed by strong electron attachment},
journal = {Plasma Sources Sci. Technol.}
}

@article{SF6111,
doi = {10.1088/0022-3727/19/12/016},
year = {1986},
month = {dec},
publisher = {},
volume = {19},
number = {12},
pages = {2363},
author = {I Gallimberti and  N Wiegart},
title = {Streamer and leader formation in SF6 and SF6 mixtures under positive impulse conditions. II. Streamer to leader transition},
journal = {J. Phys. D: Appl. Phys.},
}

@article{14,
    author = {Dhali, Shirshak K. and Pal, Anup K.},
    title = "{Numerical simulation of streamers in SF6}",
    journal = {J. Appl. Phys.},
    volume = {63},
    number = {5},
    pages = {1355-1362},
    year = {1988},
    month = {03},
    issn = {0021-8979},
    doi = {10.1063/1.339963},
}

@article{ GDYJ200807009,
author = { X Wang and F Wang and Y Qiu },
title = {均匀场中SF_6二维流注放电模型的动态仿真},
journal = {High Voltage Engineering},
volume = {},
number = {07},
pages = {1358-1362},
year = {2008},
issn = {1003-6520},
doi = {10.13336/j.1003-6520.hve.2008.07.002}
}

@article{ GDYJ201410015,
author = { X Li and  S Lin and  J Xu and  Z Li },
title = {基于FEM-FCT法的均匀电场下SF_6气体击穿特性},
journal = {High Voltage Engineering},
volume = {40},
number = {10},
pages = {3046-3053},
year = {2014},
issn = {1003-6520},
doi = {10.13336/j.1003-6520.hve.2014.10.015}
}

@ARTICLE{EN3602,
  author={Morrow, R.},
  journal={IEEE Trans. Plasma Sci.}, 
  title={A Survey of the Electron and Ion Transport Properties of SF6}, 
  year={1986},
  volume={14},
  number={3},
  pages={234-239},
  doi={10.1109/TPS.1986.4316534}}

@article{15,
doi = {10.1088/1361-6595/aae706},
url = {https://dx.doi.org/10.1088/1361-6595/aae706},
year = {2018},
month = {nov},
publisher = {IOP Publishing},
volume = {27},
number = {11},
pages = {115001},
author = {Qingqing Gao and Chunping Niu and Kazimierz Adamiak and Aijun Yang and Mingzhe Rong and Xiaohua Wang},
title = {Numerical simulation of negative point-plane corona discharge mechanism in SF6 gas},
journal = {Plasma Sources Sci. Technol.},
}

@article{zhu2021simulation,
  title={Simulation of ionization-wave discharges: a direct comparison between the fluid model and E-FISH measurements},
  author={Zhu, Yifei and Chen, Xiancong and Wu, Yun and Hao, Jinbo and Ma, Xiaoguang and Lu, Pengfei and Tardiveau, Pierre},
  journal={Plasma Sources Sci. Technol.},
  volume={30},
  number={7},
  pages={075025},
  year={2021},
  publisher={IOP Publishing},
  doi = {10.1088/1361-6595/ac0714},
  url = {https://dx.doi.org/10.1088/1361-6595/ac0714},
}

@article{10.1063/5.0131780,
    author = {Levko, Dmitry and Raja, Laxminarayan L.},
    title = "{Computational analysis of electrical breakdown of SF6/N2 mixtures}",
    journal = {J. Appl. Phys.},
    volume = {133},
    number = {5},
    pages = {053301},
    year = {2023},
    month = {02},
    issn = {0021-8979},
    doi = {10.1063/5.0131780},
    url = {https://doi.org/10.1063/5.0131780},
}

@article{10.1063/5.0006140,
    author = {Ou, Xuefeng and Wang, Lijun and Liu, Jie and Lin, Xin},
    title = "{Numerical simulation of streamer discharge development processes with multi-component SF6 mixed gas}",
    journal = {Phys. Plasmas},
    volume = {27},
    number = {7},
    pages = {073504},
    year = {2020},
    month = {07},
    issn = {1070-664X},
    doi = {10.1063/5.0006140},
    url = {https://doi.org/10.1063/5.0006140},
}

@article{ZK,
  url = {https://ui.adsabs.harvard.edu/abs/1982HTemS..20..357Z},    
author = {Zhelezniak, M B and Mnatsakanian, A K and Sizykh, S V E},
    title = "{Photoionization of nitrogen and oxygen mixtures by radiation from a gas discharge. }",
    journal = {High Temp. Sci.},
    volume = {20},
    number = {3},
    pages = {357-362},
    year = {1982}
}

@article{Bourdon_2007,
doi = {10.1088/0963-0252/16/3/026},
url = {https://dx.doi.org/10.1088/0963-0252/16/3/026},
year = {2007},
month = {aug},
publisher = {},
volume = {16},
number = {3},
pages = {656},
author = {A Bourdon and V P Pasko and N Y Liu and S Célestin and P Ségur and E Marode},
title = {Efficient models for photoionization produced by non-thermal gas discharges in air based on radiative transfer and the Helmholtz equations},
journal = {Plasma Sources Sci. Technol.},
}

@article{10.1063/1.2435934,
    author = {Luque, Alejandro and Ebert, Ute and Montijn, Carolynne and Hundsdorfer, Willem},
    title = "{Photoionization in negative streamers: Fast computations and two propagation modes}",
    journal = {Appl. Phys. Lett.},
    volume = {90},
    number = {8},
    pages = {081501},
    year = {2007},
    month = {02},
    issn = {0003-6951},
    doi = {10.1063/1.2435934},
    url = {https://doi.org/10.1063/1.2435934},
}

@article{Bujotzek_2015,
doi = {10.1088/0022-3727/48/24/245201},
url = {https://dx.doi.org/10.1088/0022-3727/48/24/245201},
year = {2015},
month = {may},
publisher = {IOP Publishing},
volume = {48},
number = {24},
pages = {245201},
author = {M Bujotzek and M Seeger and F Schmidt and M Koch and C Franck},
title = {Experimental investigation of streamer radius and length in SF6},
journal = {J. Phys. D: Appl. Phys.},
}

@article{PhysRevA.37.4396,
  title = {Formation and propagation of streamers in ${\mathrm{N}}_{2}$ and ${\mathrm{N}}_{2}$-${\mathrm{SF}}_{6}$ mixtures},
  author = {Wu, C. and Kunhardt, E. E.},
  journal = {Phys. Rev. A},
  volume = {37},
  issue = {11},
  pages = {4396--4406},
  numpages = {0},
  year = {1988},
  month = {Jun},
  publisher = {American Physical Society},
  doi = {10.1103/PhysRevA.37.4396},
  url = {https://link.aps.org/doi/10.1103/PhysRevA.37.4396}
}

@article{10.1063/5.0186055,
    author = {Levko, Dmitry and Thiruppathiraj, Sudharshanaraj and Raja, Laxminarayan L.},
    title = "{Computational analysis of the anode-directed streamers propagation in atmospheric pressure C4F7N/N2 mixtures}",
    journal = {J. Appl. Phys.},
    volume = {135},
    number = {17},
    pages = {173301},
    year = {2024},
    month = {05},
    issn = {0021-8979},
    doi = {10.1063/5.0186055},
    url = {https://doi.org/10.1063/5.0186055},
}

@article{Luque_2008,
doi = {10.1088/0022-3727/41/23/234005},
url = {https://dx.doi.org/10.1088/0022-3727/41/23/234005},
year = {2008},
month = {nov},
publisher = {},
volume = {41},
number = {23},
pages = {234005},
author = {Alejandro Luque and Valeria Ratushnaya and Ute Ebert},
title = {Positive and negative streamers in ambient air: modelling evolution and velocities},
journal = {J. Phys. D: Appl. Phys.},
}

@article{10.1063/1.1288407,
    author = {Christophorou, L. G. and Olthoff, J. K.},
    title = "{Electron Interactions With SF6}",
    journal = {J. Phys. Chem. Ref. Data},
    volume = {29},
    number = {3},
    pages = {267-330},
    year = {2000},
    month = {05},
    issn = {0047-2689},
    doi = {10.1063/1.1288407},
    url = {https://doi.org/10.1063/1.1288407},
}

@ARTICLE{8928273,
  author={Wang, Jian and Hu, Qi and Chang, Yanan and Wang, Jingrui and Liang, Ruixue and Tu, Youping and Li, Chuanyang and Li, Qingmin},
  journal={CSEE J. Power Energy Syst.}, 
  title={Metal particle contamination in gas-insulated switchgears/gas-insulated transmission lines}, 
  year={2021},
  volume={7},
  number={5},
  pages={1011-1025},
  doi={10.17775/CSEEJPES.2019.01240}}

@article{Zhao_2022,
doi = {10.1088/1361-6595/ac7ee3},
url = {https://dx.doi.org/10.1088/1361-6595/ac7ee3},
year = {2022},
month = {jul},
publisher = {IOP Publishing},
volume = {31},
number = {7},
pages = {075006},
author = {Zheng Zhao and Zongze Huang and Xinlei Zheng and Chenjie Li and Anbang Sun and Jiangtao Li},
title = {Evolutions of repetitively pulsed positive streamer discharge in electronegative gas mixtures at high pressure},
journal = {Plasma Sources Sci. Technol.},

}

@article{https://doi.org/10.1049/hve2.12119,
author = {Zhao, Zheng and Dai, Zhifeng and Sun, Anbang and Li, Jiangtao},
title = {Streamer-to-precursor transition in N2–SF6 mixtures under positive repetitive submicrosecond pulses},
journal = {High Volt.},
volume = {7},
number = {2},
pages = {382-389},
doi = {https://doi.org/10.1049/hve2.12119},
url = {https://ietresearch.onlinelibrary.wiley.com/doi/abs/10.1049/hve2.12119},
year = {2022}
}

@article{PhysRevA.35.1778,
  title = {Properties of streamers and streamer channels in ${\mathrm{SF}}_{6}$},
  author = {Morrow, R.},
  journal = {Phys. Rev. A},
  volume = {35},
  issue = {4},
  pages = {1778--1785},
  numpages = {0},
  year = {1987},
  month = {Feb},
  publisher = {American Physical Society},
  doi = {10.1103/PhysRevA.35.1778},
  url = {https://link.aps.org/doi/10.1103/PhysRevA.35.1778}
}

@article{10.1063/5.0223522,
    author = {Feng, Zihao and Jiang, Yuanyuan and Zhang, Liyang and Liu, Zhigang and Wang, Kai and Wang, Xinxin and Zou, Xiaobing and Luo, Haiyun and Fu, Yangyang},
    title = "{Microscopic characteristics of SF6 partial discharge induced by a floating linear metal particle}",
    journal = {Appl. Phys. Lett.},
    volume = {125},
    number = {13},
    pages = {134101},
    year = {2024},
    month = {09},
    issn = {0003-6951},
    doi = {10.1063/5.0223522},
    url = {https://doi.org/10.1063/5.0223522},
}

\end{document}